\documentclass[twocolumn]{aastex6}
\usepackage{epsfig}
\usepackage{natbib}
\usepackage{amsmath}
\usepackage{rotating}
\newcommand{\kms}{km~s$^{-1}$}
\newcommand{\Msun}{M_{\odot}}

\newcommand{\kmsMpc}{km~s$^{-1}$~Mpc$^{-1}$}

\begin{document}
\title{Cosmicflows-3}

\author{R. Brent Tully,}
\affil{Institute for Astronomy, University of Hawaii, 2680 Woodlawn Drive,
 Honolulu, HI 96822, USA}
\and
\author{H\'el\`ene M. Courtois}
\affil{Universit\'e Claude Bernard Lyon I, Institut de Physique Nucl\'eaire, Lyon, France}
\and
\author{Jenny G. Sorce}
\affil{Leibniz-Institut f\"ur Astrophysik, D-14482 Potsdam, Germany}

\begin{abstract}
The {\it Cosmicflows} database of galaxy distances that in the 2nd edition contained 8,188 entries is now expanded to 17,669 entries.  The major additions are 2,257 distances that we have derived from the correlation between galaxy rotation and luminosity with photometry at $3.6~\mu$m obtained with {\it Spitzer Space Telescope} and 8,885 distances based on the Fundamental Plane methodology from the 6dFGS collaboration.  There are minor augmentations to the Tip of the Red Giant Branch and Type I$a$ supernova compilations.  A zero point calibration of the supernova luminosities give a value for the Hubble Constant of $76.2 \pm 3.4 \pm 2.7$ ($\pm$ rand. $\pm$ sys.)~\kmsMpc.  Alternatively, a restriction on the peculiar velocity monopole term representing global infall/outflow implies $H_0 = 75 \pm 2$~\kmsMpc.

\bigskip
\end{abstract}

\smallskip
\section{Introduction}

{\it Cosmicflows-3} is a compendium of galaxy distance that builds on two earlier releases \citep{2008ApJ...676..184T, 2013AJ....146...86T} and draws on both original material and information from the literature.  The most important original material in {\it Cosmicflows-3} extends the correlation between galaxy rotation and luminosity, hereafter referred to as TF \citep{1977A&A....54..661T}, by using infrared photometry obtained with {\it Spitzer Space Telescope}.  The most important addition from the literature is the extensive Fundamental Plane (FP) sample derived from the Six Degree Field Galaxy Survey (6dFGS) of the southern celestial hemisphere \citep{2014MNRAS.445.2677S}.  Less substantial additions include new distances based on identification of the Tip of the Red Giant Branch (TRGB) in {\it Hubble Space Telescope} (HST) images  and an update on literature distance determinations from Type I$a$ supernova (SNI$a$) observations \citep{2014ApJ...795...44R, 2015ApJS..219...13W}.
 
The discussion will begin by describing the new {\it Spitzer} sample.  The study involves a review of the calibration procedure, followed by application to all relevant galaxies with {\it Spitzer} photometry.    Following that, attention will be given to the integration of 6dFGS  FP distances, heedful of the need to maintain a constant zero point scale. The next topic will visit the status of TRGB observations. Finally, similar zero-point considerations as with 6dFGS FP will guide the acceptance of SNI$a$ information into the assembly.  The SNI$a$ calibration will be used to infer a value for the Hubble Constant at redshifts $0.05 < z < 0.6$, beyond the domain of velocity distortions.

Use will be made of a new group catalog \citep{2015AJ....149..171T}.  Associations with groups enable comparisons within and between methodologies.  Large scale flow studies are improved by group averaging distances and velocities.

\section{Luminosity--Linewidth Distances with Spitzer Photometry}

The viability of using {\it Spitzer Space Telescope} photometry in a band at $3.6~\mu$m, [3.6], to represent luminosities in the TF relation was demonstrated by \citet{2013ApJ...765...94S}.  Photometry at $3.4~\mu$m with $WISE$, {\it Wide-field Infrared Satellite Explorer}, is similarly useful \citep{2013ApJ...771...88L, 2014ApJ...792..129N}.
Obvious advantages of the infrared are virtual elimination of obscuration concerns and a matching to the spectral output of the old stars that dominate the baryon mass budget.  Obvious advantages of satellite observations are vastly reduced backgrounds and photometric integrity over the entire sky.  A slight disadvantage compared with the familiar $I$ band turns out to be the introduction of a color term, but the color correction required to minimize dispersion is small.

Observations with {\it Spitzer Space Telescope} are targeted.  There have been two major programs of relevance to the current study.  The first of these was $S^4G$, the {\it Spitzer Survey of Stellar Structure in Galaxies} \citep{2010PASP..122.1397S} which acquired imaging photometry in the 3.6 and 4.5 micron bands, [3.6] and [4.5], for 2352 disk galaxies at $\vert b \vert>30^{\circ}$, and with distance, magnitude and size limits.  The intent of the $S^4G$ program was to study the structural properties of nearby disk galaxies \citep{2010PASP..122.1397S, 2015ApJS..219....4S} but a substantial fraction of the $S^4G$ sample is also of interest for this work.  The second major program was our $CFS$, {\it Cosmic Flows with Spitzer} \citep{2014MNRAS.444..527S}.  Only galaxies of interest for the acquisition of distances were considered in this program and observations were restricted to the 3.6 $\mu$m band.  In $CFS$, a primary selection criterion was low galactic latitude, $\vert b \vert < 30^{\circ}$, both as a complement to $S^4G$ and particularly because it was appreciated that it is in crowded stellar fields that {\it Spitzer} has the greatest advantage over the concurrent {\it WISE} because of superior spatial resolution.  The $CFS$ observations at higher latitudes favored extreme edge-on systems \citep{1999BSAO...47....5K}, choosing among those that already have adequately observed HI line profiles.

The {\it Spitzer} sample encompasses observations obtained in a number of earlier programs.  The relevant images are available at the {\it Spitzer} archive.  In the interest of maintaining homogeneity, all the $CFS$ and non-$S^4G$ archival images were analyzed with the Archangel photometry package \citep{2007astro.ph..3646S} as discussed by \citet{2012AJ....144..133S, 2014MNRAS.444..527S}.  The integration of Archangel and $S^4G$ photometry will be discussed in the next Section.  In total, 2257 galaxy distances could be determined based on {\it Spitzer} photometry.

\subsection{3.6 Micron Magnitude vs. Linewidth Calibration}  

The calibration of the correlation between [3.6] magnitudes and HI profile widths follows previously described procedures for the calibration at $I$ band \citep{2012ApJ...749...78T}, the {\it WISE} $W1$ and $W2$ bands \citep{2014ApJ...792..129N}, and the {\it Spitzer} [3.6] band \citep{2013ApJ...765...94S, 2014MNRAS.444..527S}.  The slope calibration averages over separate samples that are each approximately volume limited to specified magnitude limits and that are representative of a range of galaxy densities and dominant types.

Thirteen clusters described earlier \citep{2000ApJ...533..744T, 2012ApJ...749...78T, 2014ApJ...792..129N} provide the slope calibration.  Initially, linear fits are made to the separate samples within each cluster, with errors taken in linewidths; the so-called inverse TF relation.  Then through an iterative procedure the 13 clusters are shifted in the magnitude domain to overlay on the Virgo sample with a best fit slope to the ensemble and minimized rms dispersion.

The absolute zero point is set by galaxies with distances determined by the Cepheid period-luminosity relation or the TRGB method.  The Population I Cepheid distances are set by an assumed modulus for the Large Magellanic Cloud of 18.48 \citep{2012ApJ...758...24F}.  The Population II TRGB distances have been demonstrated to be in good agreement with the Cepheid scale \citep{2007ApJ...661..815R}.  Both Cepheid and TRGB methods give distances to NGC~4258 that agree with the maser distance  \citep{2013ApJ...775...13H}.  It is assumed that the absolute zero point calibrators obey the same relationship as demonstrated by the 13 cluster template; ie, they randomly sample a relationship with the same slope.  Minimization of the rms dispersion of the zero point calibrators with this fixed slope gives the absolute [3.6] vs. log linewidth calibration.

The important raw input, besides the absolute calibrator distances, are total [3.6] magnitudes, the linewidth estimator $W_{mx}$ \citep{2009AJ....138.1938C, 2011MNRAS.414.2005C}, and inclinations.  Minor corrections are made for reddening and velocity dependent effects.  Insofar as the 13 cluster and zero point calibration are concerned there are only mild updates to the [3.6] photometry as reported by \citet{2014MNRAS.444..527S} and the HI linewidth values used by \citet{2014ApJ...792..129N}.    There were 291 cluster                                                                                                                                                                                                                                                                                                                                                                    calibrators available to Neill et al. with the all-sky {\it WISE} coverage while only a subset of 285 are available for the {\it Spitzer} calibration because a few calibrators have not been targeted in any {\it Spitzer} program.  As a consequence of minor group membership revisions discussed in Section~\ref{sec:template}, 305 galaxies now contribute to the 13 cluster template.

Particular attention has been given to the issue of inclinations with the present re-calibration.  Uncertain inclinations are a dominant source of error.  At $3.6~\mu$m, uncertainties in reddening are minor.  The larger uncertainties are projection corrections to linewidths.  Historically, inclinations are derived from projected minor to major axis ratios assuming galaxies appear circular viewed face-on and have a specified thickness.  Here, an edge-on disk system is assumed to have the ratio $b/a=0.20$, with the rational for that choice discussed by \citet{2000ApJ...533..744T}.

Unfortunately, most sources of axial ratios in the literature are unreliable.  Measures based on infrared photometry have proven to be un-useful, whether from {\it Spitzer}, {\it WISE}, or {\it 2MASS}, measures by our team included.  Bulges and bars are sources of confusion.  In the minority of cases observed with long integrations by {\it GALEX} \citep{2007ApJS..173..682M}, the resultant ultraviolet images provide reasonable isolation of disks with the virtual elimination of bulges and bars.  Generally the fallback for inclination estimates is optical images, with preference for the most blueward band available.

Too often, the derivation of inclinations is not the primary concern of a photometric program, with the consequence that reported axial ratios do not realistically translate to inclinations.  Inclinations were a concern of the photometry by the Cornell group and generally their axial ratios give good inclinations \citep{1997AJ....113...22G, 2006ApJ...653..861M, 2007ApJS..172..599S}.  Lyon Extragalactic Database (LEDA) axial ratios are decent for inclinations if, but only if, the source is \citet{1996A&A...311...12P}.   The {\it Cosmicflows-2} compilation uses these sources as well as measurements by our collaboration.   The agreement between the most reliable sources reflect uncertainties at the level of $\pm 4^{\circ}$.

Visual inspection can reveal instances where axial ratio measurements are misleading.  On the side toward higher inclinations, galaxies manifestly viewed almost edge-on can have misrepresentative axial ratios because of bulges.  These cases are generally not difficult to correct.  On the side toward face-on inclinations, high surface brightness bars embedded in low surface brightness disks cause problems.  Axial ratios may describe bars although the disks, if properly identified, may suggest more face-on aspects.

Ultimately, one is faced with the probability that some disk galaxies are poorly approximated by oblate spheroids.  Tidal distortions, warps, strong bars, spiral features, and pronounced surface brightness transitions create confusion.  With each of these problems, the manifestation to the observer depends on relative position angle and the inclination, the parameter of primary interest.  From experience, roughly a third of candidates pose significant difficulties.

The strategy employed in the current study is to begin with quantitative measurements of axial ratios from photometry, comparing and often averaging over different sources \citep{2013AJ....146...86T}, but then making visual evaluations.  In most cases the averaged quantitative axial ratio from sources deemed reliable is considered to provide a good representation of the inclination.  In cases that are considered problematic, inclinations and equivalent axial ratios are estimated visually by comparison with a trainer set of examples with reasonably established inclinations.

Our strategy is only partially satisfactory.  The available imaging material is heterogeneous.  Deep {\it GALEX} ultraviolet images are excellent but in limited supply.  {\it Spitzer} infrared images are always available and sufficiently sensitive but have unsatisfactory bulge and disk contamination.  In the near future, digital area photometry of the entire sky at blueward optical bands will be available from Pan-STARRS and SkyMapper observations \citep{2010SPIE.7733E..0EK}.  Access to that material in a future {\it Cosmicflows} release will be appreciated because the qualitative aspects of the present procedure could produce resolution, hence distance dependent biases.

\subsection{Thirteen Cluster Slope Template}
\label{sec:template}

\begin{figure}[]
\includegraphics[scale=.4]{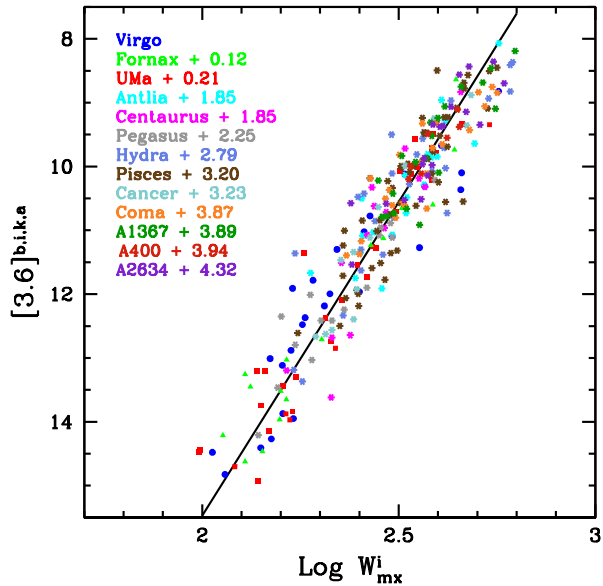}
\caption{[3.6] vs. HI linewidth template using samples drawn from 13 clusters and shifted to a best fit at the relative distance of the Virgo Cluster.  Colors distinguish the 13 separate clusters. The rms scatter is 0.55 mag.}
\label{tfsatvir}
\end{figure}

The individual fits to 13 clusters are not very different from previous fits shown in \citet{2014MNRAS.444..527S}.  The {\it Spitzer} photometry and HI linewidth information is essentially the same.  A new study of galaxy groups \citep{2015AJ....149..171T} has lead to some modifications.  Spatial and velocity associations between the $Spitzer$ sample and the 13 calibration clusters have lead to the identification of 26 additional cluster members with appropriate properties.  The group analysis resulted in a split of the Abell 1367 Cluster into two components (nest 100005 is A1367 and nest 120005 is  displaced by $2^{\circ}$ and lies slightly closer).  The new calibration uses 17 galaxies associated with the revised A1367.  In all, the new calibration uses 305 galaxies, an increase from 287 in \citet{2014MNRAS.444..527S}.  In the particular case of the Virgo Cluster, we continue to only use galaxies across a restricted spatial domain that should minimize background contamination.  This issue is re-evaluated in Section~\ref{virgo} once the calibration is completed and applied to the full $Spitzer$ sample.  In this new work, the greatest cause for changes comes from revised inclinations.  Figure~\ref{tfsatvir} shows the new $Spitzer$ [3.6] vs. linewidth correlation for 13 clusters shifted to an optimal fit at the distance of the Virgo Cluster. 

\begin{figure}[]
\includegraphics[scale=.4]{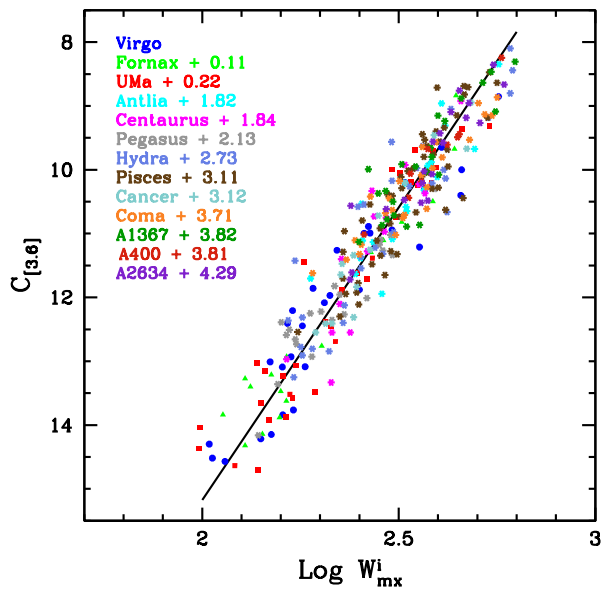}
\caption{Color corrected version of the [3.6] vs. HI linewidth template. Scatter 0.48 mag.}
\label{tfcatvir}
\end{figure}

The infrared TF correlation has a color term \citep{2013ApJ...765...94S, 2014MNRAS.444..527S, 2014ApJ...792..129N}.  An adjustment that reduces scatter can be made if the $I-[3.6]$ color is available.  The color adjustment is only slightly different from the earlier formulation.  Now:
\begin{equation}
\Delta[3.6] = -0.55 (I - [3.6] +1) .
\end{equation}  
Application of the color term subtly but significantly improves the infrared TF correlation.  The color corrected version of Figure~\ref{tfsatvir} is shown as Figure~\ref{tfcatvir}.  Here, for the 15 calibrator galaxies lacking $I$ magnitudes the color adjustment is taken to be zero.  The slope without color adjustment is $-9.72 \pm 0.19$ and after color adjustment is $-9.16 \pm 0.17$.

As noted in Sorce et al. and Neill et al. there is a small selection bias that becomes increasingly important at distances where the faintest galaxies in a sample have intrinsic magnitudes near $M^{\star}$, characterizing the exponential cutoff of the Schechter function \citep{1976ApJ...203..297S}.  The current sample closely approximates the Neill et al. sample and the bias adjustments advocated in that paper are incorporated in this work.

\subsection{Zero Point Calibration} 

The absolute scale of the TF relations is set by nearby galaxies with alternate distance estimates that are considered the best available.  Here, 33 of the 37 galaxies considered by \citet{2014ApJ...792..129N} provide the calibration.  Three of the Neill et al. calibrators lack {\it Spitzer} photometry (M33, NGC~3109, NGC~4945) while one (NGC~4535) is given  a revised inclination that fails our $45^{\circ}$ limit.  The input distances are derived in 25 cases from the Cepheid period-luminosity relation assuming a fiducial distance modulus for the Large Magellanic Cloud of 18.48 \citep{2012ApJ...758...24F, 2014AJ....147..122D}, in 19 cases from the TRGB method assuming the calibration by \citet{2007ApJ...661..815R}, and in the special case of NGC 4258 from the geometric model of the maser emission  \citep{2013ApJ...775...13H}.  The calibration of the Population I Cepheids and Population II red giant branch tip are independent, yet give distances that agree at the level of 0.01 mag. Eleven galaxies have both Cepheid and TRGB measures.  The Cepheid, TRGB, and maser methodologies agree on the distance to NGC~4258 of $7.57\pm0.10$~Mpc.

The sample of galaxies with well established distances does not approximate a volume limited sample so should not be used to define the slope of the dependency between rotation rate and luminosity.  The slope is defined by the 13 cluster template.  Assuming that slope, the known absolute luminosity and inclination corrected line width of each of the 33 absolute calibrators provides an independent zero point estimation.  The rms minimized deviation for the 33 galaxies provides the best fit solution.

This procedure was carried out separately for the basic [3.6] vs. linewidth correlation and for the color adjusted version.  The calibrations are
\begin{equation}
M_{3.6}^{b,i,k,a} = -20.35\pm0.09 - 9.72\pm0.19 ({\rm log} W_{mx}^i -2.5)
\label{eq:uc}
\end{equation}
\begin{equation}
M_C = -20.38\pm0.08 - 9.16\pm0.17 ({\rm log} W_{mx}^i -2.5)
\label{eq:cc}
\end{equation}
where $W_{mx}^i$ are inclination corrected linewidths, $M_{3.6}^{b,i,k,a}$ are absolute magnitudes in the {\it Spitzer} [3.6] band corrected for absorption within our Galaxy and the host galaxy, for translation of the rest frame with respect to the filter response and an aperture correction \citep{2013ApJ...765...94S}, and $M_C$ is the color adjusted modification of $M_{3.6}^{b,i,k,a}$.  The rms dispersions of the absolute calibrators from these relations are $\pm 0.54$ for the basic correlation and $\pm 0.45$ for the color adjusted version.

\begin{figure}[!]
\includegraphics[scale=.4]{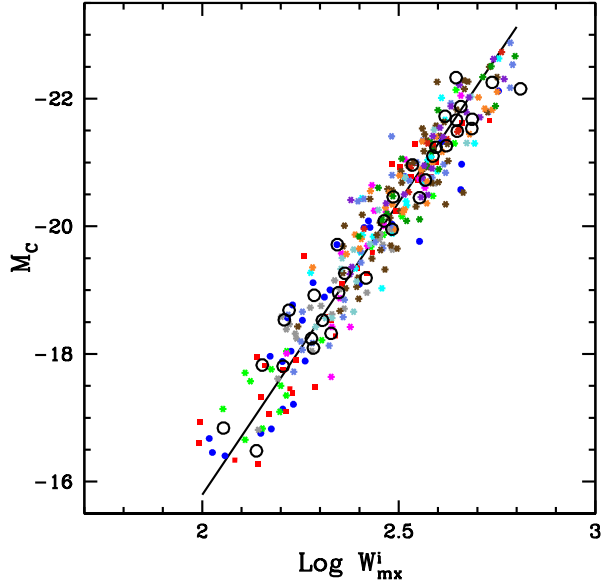}
\caption{Color corrected absolute [3.6] luminosity vs. HI linewidth correlation.  Absolute luminosity scale is set by 33 galaxies identified by large open circles.  Zero-point scatter is 0.45 mag.}
\label{tfc}
\end{figure}

Statistics on the cluster fits are assembled in Table~\ref{tbl:13clust}.  Distances are provided alternatively with, and without, color corrections.  The differences are not statistically significant but uncertainties are slightly reduced through the color adjustment and color adjusted distances are preferred.  The table also includes the cluster distances determined in the {\it WISE} $W1$ band at $3.4~\mu$m by \citet{2014ApJ...792..129N}, our most recent previous calibration.  The Neill et al. cluster distances are on average 3\% greater.  The newly added galaxies to the cluster template sample have a slight tendency to lie to the low linewidth side of the mean TF relation on average, something taken to be a statistical vagary but which lowers distances and increases the calculated $H_0$.  The difference from the Neill et al. calibration is less than the $1 \sigma$ error on the color corrected zero point, Eq~\ref{eq:cc}. 

\begin{deluxetable*}{lccccccrcrcrr}
\tablenum{1}
\tablecaption{Template Cluster Distances}
\label{tbl:13clust}
\tablewidth{0in}
\tablehead{\colhead{Cluster} & \colhead{$\mu_{cc}$} & \colhead{$\pm$} & \colhead{$b$} & \colhead{$\mu_{uc}$} & \colhead{$\pm$} & \colhead{$V_{mod}$} & \colhead{$d_{cc}$} & \colhead{err} & \colhead{$d_{uc}$} & \colhead{err} & \colhead{$d_{W1}$} & \colhead{$H_i$}}
\startdata
Virgo     & 31.029 & 0.136 & 0.000 & 30.908 & 0.164 &  1495 &  16.1 &  1.0 &  15.2 &  1.2 &  16.2 &  93.1 \\
UrsaMajor & 31.182 & 0.120 & 0.000 & 31.123 & 0.138 &  1079 &  17.2 &  1.0 &  16.8 &  1.1 &  17.2 &  62.6 \\
Fornax    & 31.099 & 0.148 & 0.000 & 31.029 & 0.169 &  1358 &  16.6 &  1.2 &  16.1 &  1.3 &  17.5 &  81.9 \\
Antlia    & 32.835 & 0.140 & 0.040 & 32.800 & 0.154 &  3198 &  36.9 &  2.4 &  36.3 &  2.7 &  39.0 &  86.7 \\
Centaurus & 32.808 & 0.168 & 0.000 & 32.756 & 0.191 &  3823 &  36.4 &  2.9 &  35.6 &  3.3 &  38.5 & 104.9 \\
Pegasus   & 33.211 & 0.132 & 0.000 & 33.162 & 0.148 &  3062 &  43.9 &  2.8 &  42.9 &  3.0 &  44.0 &  69.8 \\
Hydra     & 33.714 & 0.154 & 0.010 & 33.711 & 0.176 &  4088 &  55.3 &  4.1 &  55.2 &  4.7 &  62.1 &  73.9 \\
Pisces    & 34.105 & 0.104 & 0.020 & 34.130 & 0.122 &  4759 &  66.2 &  3.3 &  67.0 &  3.9 &  68.3 &  71.9 \\
Cancer    & 34.149 & 0.125 & 0.020 & 34.155 & 0.148 &  5059 &  67.6 &  4.0 &  67.8 &  4.8 &  64.2 &  74.9 \\
Coma      & 34.792 & 0.118 & 0.040 & 34.818 & 0.141 &  7370 &  90.9 &  5.1 &  92.0 &  6.2 &  93.9 &  81.1 \\
A1367     & 34.863 & 0.121 & 0.080 & 34.879 & 0.143 &  6969 &  93.9 &  5.4 &  94.6 &  6.4 &  98.1 &  74.2 \\
A400      & 34.950 & 0.115 & 0.110 & 34.957 & 0.134 &  7228 &  97.7 &  5.3 &  98.0 &  6.2 & 100.0 &  74.0 \\
A2634/66  & 35.347 & 0.123 & 0.079 & 35.309 & 0.143 &  8938 & 117.3 &  6.8 & 115.3 &  7.9 & 116.9 &  76.2 \\
\enddata
\tablecomments{$\mu_{cc}$ and $\mu_{uc}$ are color corrected and uncorrected distance moduli respectively (mags); $d_{cc}$ and $d_{uc}$ are corresponding distances (Mpc); $d_{W1}$ is distance (Mpc) from \citet{2014ApJ...792..129N}; $V_{mod}$ (\kms) and $H_i$ (\kmsMpc) are defined in Eq.~\ref{hi}. }
\end{deluxetable*}

\subsection{Field Sample}

Distances are estimated for 2257 galaxies.  These galaxies meet the following criteria: (i) measured linewidths with uncertainties $\le 20$~\kms, (ii) satisfactory {\it Spitzer} [3.6] photometry, (iii) morphological types Sa and later, (iv) inclinations more edge-on than $45^{\circ}$, (v) HI not confused by an adjacent system, and (vi) not morphologically disrupted or peculiar.  The last criterion is subjective; we tend to be inclusive.

Distances require correction for a minor selection bias.  The bias is minimized by using the inverse TF relation with errors taken in linewidth \citep{1994ApJS...92....1W} but a small bias persists because there are more faint galaxies available to scatter brightward than bright galaxies to scatter faintward.  The present cluster calibration samples are almost identical to those used in the \citet{2014ApJ...792..129N} study in the {\it WISE} $3.4~\mu$m band and the bias correction is taken to be the same.  
\begin{equation}
b = 0.004 (\mu-31)^{2.3}
\end{equation}
where $b$ is the bias correction and $\mu$ is a measured distance modulus.  For $\mu \le 31$, $b=0.$

Evaluation of individual Hubble parameters, $H_i = V_i/d_i$, where $V_i$ and $d_i$ are measured velocities and distances, provides a way of identifying aberrant measurements.  Here, the velocity in the Hubble parameter is in the CMB frame modified by cosmological curvature terms
\begin{equation}
\begin{split}
H_i = V_{mod}^i/d_i = \\ 
{{cz_i}\over{d_i}} {\{1 + {{1}\over{2}} [1 - q_0]z - {{1}\over{6}} [1 - q_0 -3 q_0^2 + j_0]z^2\}}
\end{split}
\label{hi}
\end{equation}
where $z_i$ is redshift, the jerk parameter $j_0 = 1$ and the acceleration parameter $q_0 = {{1}\over{2}} ( \Omega_m -2 \Omega_{\Lambda} ) = -0.595$ (taking $\Omega_m=0.27$, $\Omega_m+\Omega_{\Lambda}=1$).

\begin{figure}[t]
\includegraphics[scale=.4]{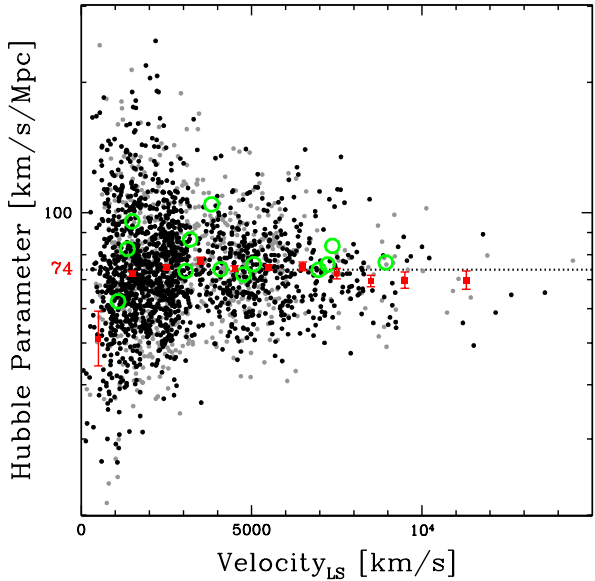}
\caption{Hubble parameter $H_i=V_{mod}^i/d_i$ for the full {\it Spitzer} sample.  Distances from the color adjusted relation are black while distances from the basic correlation (ie, lacking $I$ band magnitudes for the color adjustment) are in grey.  Averages of $H_i$ in 1000~\kms\ intervals are plotted in red with 1 standard deviation errors.  The average for 647 values of $H_i$ with $V_{LS} > 4000$~\kms\ is 73.6~\kmsMpc.  Large green circles identify the Hubble parameter values for the 13 template clusters.}
\label{Hi}
\end{figure}

Cases with $H_i$ that deviate by $>3\sigma$ from the mean were examined.  In a majority of such cases, the target could be seen in retrospect to fail our selection criteria and could be removed from the sample.  However there were cases with no evident anomaly.  Indeed, it has been long known that about 3\% of candidates have excursions greater than $3\sigma$ from the TF relation, some reaching $4-5\sigma$, statistically improbable if the errors are Gaussian.  The situation is the same with the current sample.  It is to be appreciated that large deviations from the mean can be physically realistic for very nearby systems where true peculiar velocities can dominate over errors.  After due consideration for this possibility, galaxies with deviations in $H_i$ greater than $3.7\sigma$ are rejected.  Figure~\ref{Hi} shows the distribution of $H_i$ with systemic velocity for the accepted sample.

It is seen that the {\it Spitzer} sample is dense at velocities less than 3000~\kms, then falls off, but picks up again through 7000~\kms, then rapidly falls off.  This behavior is a result of mixing two selection criteria.  There are many targets at $V<3000$~\kms\ as a reflection of the upper distance limit of the $S^4G$ sample and the intent of our $CFS$ program to supplement $S^4G$ at Galactic latitudes below $30^{\circ}$.  Then at higher velocities the sample is dominated by extreme edge-on galaxies from \citet{1999BSAO...47....5K}.  These galaxies have very low axial ratios, hence have small bulges, hence are typically Sc spirals.  Targets were selected that already had well observed HI profiles.  The number observed was governed by the restricted availability of {\it Spitzer} observing time.

In Figure~\ref{Hi}, distances estimated from the color adjustment formula, Eq.~\ref{eq:cc}, lead to the black points while the uncorrected distances through Eq.~\ref{eq:uc}  lead to the grey points.  Average values of the Hubble parameter, $V_{mod}^i/d_i$, in 1000~\kms\ bins are shown in red with standard deviations of the mean.  Values for the 13 template calibrator clusters are shown by large open green circles.  Large excursions of the Hubble parameter are seen at low velocities where peculiar velocities are significant.  Averaging in the modulus over 647 galaxies with $V_{LS} > 4000$~\kms\ gives $<H_i> = 73.6 \pm 7$~\kmsMpc.

\section{Six Degree Field Galaxy Survey Distances}
\label{sec:6dfgs}

\citet{2014MNRAS.445.2677S} have published a sample of great importance for velocity field studies.  They have combined measures of the central velocity dispersion of galaxies obtained in the course of the eponymous Six Degree Field Redshift Survey \citep{2009MNRAS.399..683J} with photometry from 2MASS, the Two Micron All-Sky Survey \citep{2000AJ....119.2498J}, to derive distances from the FP correlation \citep{1987ApJ...313...59D, 1987ApJ...313...42D}.  They give distances for 8885 galaxies within 16,000~\kms, all in the south celestial hemisphere.  The accuracy for each measure is $\sim25\%$.  Details regarding the observed parameters and the fundamental plane fitting are discussed by \citet{2012MNRAS.427..245M} and \citet{2014MNRAS.443.1231C}.

The 6dFGS sample is of particular interest because coverage of the celestial south had previously been deficient.  The most sensitive radio telescopes are in the north, leading to an imbalance in coverage with the TF method.  Inclusion of the 6dFGS sample roughly doubles the number of galaxies with measured distances within $z=0.1$ and provides superior all-sky inclusiveness within 16,000~\kms.

\citet{2014MNRAS.445.2677S} provide distances as a fraction of the distance the galaxies would have if they obey the Hubble law.  The angular diameter distances that are the natural product of an FP analysis were converted to co-moving distances by Springob et al. and are further converted to luminosity distances here, consistent with our other measures: $d_L = (1+z) d_c$ where $d_L$ and $d_c$ are luminosity and co-moving distances respectively and $z$ is redshift.  It remains only to establish a zero-point match between the 6dFGS data and the other elements of our catalog.

As an initial scaling, to be roughly but not exactly consistent with the zero-point of the rest of our sample, a global expansion value of 75~\kmsMpc\ was assumed for 6dFGS.  We confirm that the scatter in distances reflected in the scatter of the individual Hubble parameter measures is the advertised 26\%.  For 112 galaxies in 26 6dFGS groups that are in common with the {\it Cosmicflows-2} compilation, there is agreement in distance moduli on average, within $1\sigma$.  The dispersions about the mean group distance moduli are as anticipated: for 8 groups with $6-10$ measurements (60 galaxies total) the 6dF rms is $\pm 0.50$, while for the same galaxies in {\it Cosmicflows-2}, the rms is $\pm 0.42$.

Here are modulus comparisons between 6dFGS at the fiducial scale $H_0=75$ and {\it Cosmicflows-2} by source of the CF2 distance.

SNI$a$ (15 cases): $<\mu_{6df}-\mu_{cf2}>=-0.03\pm0.19$ with scatter $\pm0.74$

SBF~ (35 cases): $<\mu_{6df}-\mu_{cf2}>=-0.11\pm0.08$ with scatter $\pm0.50$

FP~~ (34 cases): $<\mu_{6df}-\mu_{cf2}>=0.14\pm0.06$ with scatter $\pm0.35$

TF~~ (76 cases): $<\mu_{6df}-\mu_{cf2}>=-0.55\pm0.07$ with scatter $\pm0.57$

All but TF (84 cases): $<\mu_{6df}-\mu_{cf2}>=0.00\pm0.05$ with scatter $\pm0.48$

\begin{figure}[!]
\begin{center}
\includegraphics[scale=.4]{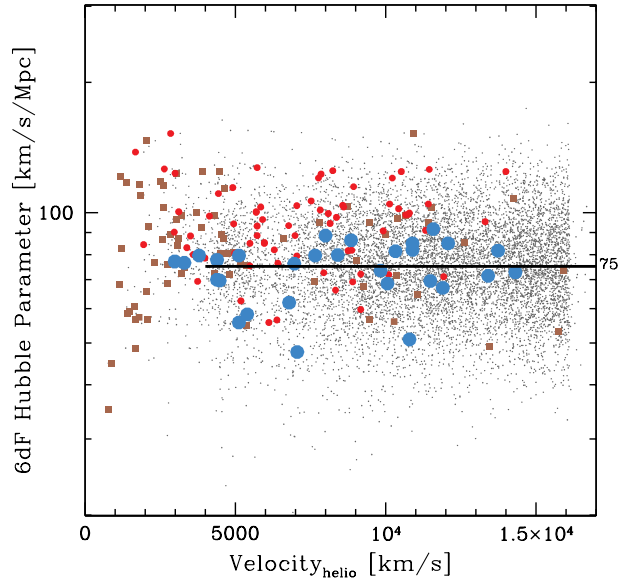}
\caption{Hubble parameter, $V_{mod}/d$, for sub-samples. The full 6dFGS sample is represented by small grey dots, with the distance scale set by the choice $H_0=75$~\kmsMpc.  The average value for galaxies with $V_{helio} > 4000$~\kms\ is 75~\kmsMpc.  The large blue points represent mean values for 29 groups with 6dFGS distances.  The brown and red points represent the values for individual galaxies within the 6dFGS sample that also have alternative distance estimates: brown squares if SNI$a$, SBF, or, FP; red circles if TF.}
\label{H6df}
\end{center}
\end{figure}

 There is statistical agreement with all but the TF comparisons.  The issue is pursued in Figure~\ref{H6df}.  The cloud of grey points represent the entire 6dFGS sample of 8885 galaxies tied to a fiducial $H_0=75$~\kmsMpc\ zero-point.  The average of the 8668 cases above 4000~\kms, the domain relatively uncontaminated by peculiar velocities, is 75~\kmsMpc.  The group comparisons (blue circles) and individual comparisons involving SNI$a$, SBF, and FP alternatives (brown squares) are consistent with the fiducial input of 75~\kmsMpc.  The red circles representing comparisons with TF lie systematically high; the 6dFGS distances appear to be too small in these cases.

The galaxies in the 6dFGS $-$ TF overlap are overwhelmingly spirals; 80\% have 6dFGS spiral morphologies and all meet the type Sa or later criterion of the TF sample.  The indication of an anomaly in the spiral component of the 6dFGS FP distances compels a closer look.  The 6dfGS collaboration provide a numerical morphological type for each of the galaxies in their sample, running from $M_t=0$ for ellipticals, through $M_t=2$ for lenticulars, to $M_t=4$ for spirals \citep{2014MNRAS.443.1231C}.  We calculate the Hubble parameter for the individual galaxies in the 6dFGS sample, $H_i=V_i/d_i$ where the velocity is a group average if possible, and then we average the Hubble parameter values in $M_t$ bins.

\begin{figure}[t]
\begin{center}
\includegraphics[scale=.4]{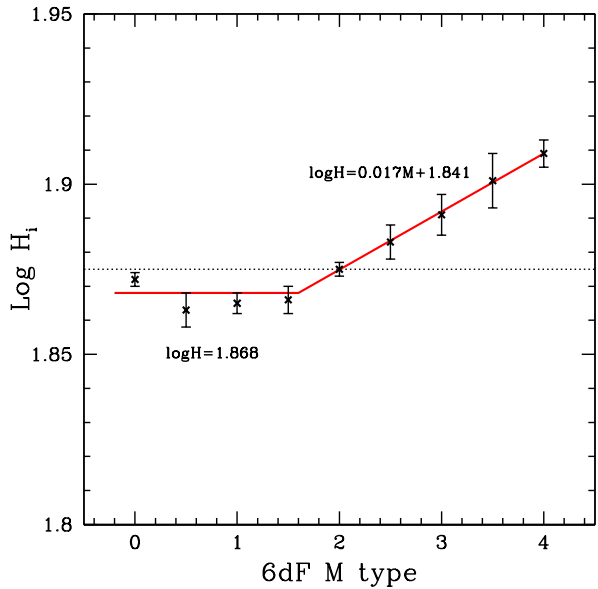}
\caption{Mean values of the Hubble parameter for 6dFGS galaxies with velocities greater than 4000~\kms\ binned by 6dFGS morphological M type.  The fiducial setting for the sample is $H_0=75$~\kmsMpc\ identified by the horizontal dotted line.  The solid red lines describe statistical deviations from the fiducial value as a function of M type.}
\label{mhm}
\end{center}
\end{figure}

The resultant means and $1\sigma$ standard deviations are displayed in Figure~\ref{mhm}.  There is an unambiguous trend for values of $H_i$ to increase with increasing $M_t$, implying distances are measured relatively too close with increasing $M_t$.  The trend is reasonably captured by the red lines in the figure, with mean $H_i$ constant at $M_t \le 1.6$ and increasing at $M_t > 1.6$.  These fits with a break at $M_t = 1.6$ translate to corrections in distance moduli.  The adjustments slightly affect the modulus comparisons with the CF2 scale discussed earlier in this section.  Minimization of the offset with the 84 galaxies with SNI$a$, SBF, or FP distances on the CF2 scale requires reducing 6dFGS galaxy moduli by 0.007 mag from the fiducial scale set by $H_0 = 75$.  This minor offset is combined with the type correction in our formulation of adjustments to 6dFGS distances to optimize the linkage to the CF2 scale:
\begin{equation}
\mu_{6df}^c - \mu_{6df}^{fid} = -0.042 + 0.085 (M_t-1.6)
\end{equation} 
if $M_t > 1.6$ and 
\begin{equation}
\mu_{6df}^c - \mu_{6df}^{fid} = -0.042 
\end{equation} 
if $M_t \le 1.6$, where $\mu_{6df}^c$ and $\mu_{6df}^{fid}$ are corrected and fiducial ($H_0=75$) distance moduli.

The FP and TF correlations have related physical origins.  The TF method is based on the assumption that rotation dominates the kinematics as is generally justified in mainly disk systems while the FP method is based on the assumption that dispersion dominates kinematics as found in early-type systems and bulges.  Bulge/disk type dependencies can be expected in FP and TF correlations.  With the Spitzer TF calibration the color term is a proxy for type.  There have been attempts to find unification correlations, linking luminosities, dimensions, and kinematics across galaxy types \citep{2011ApJ...727..116Z, 2014ApJ...795L..37C}.

\section{Tip of the Red Giant Branch Distances}

Imaging that resolves stars in nearby galaxies with {\it Hubble Space Telescope} is ongoing, both by members of our collaboration and by others.  Regardless of the source, the archival images are analyzed by our team in a coherent fashion \citep{2006AJ....132.2729M, 2007ApJ...661..815R, 2009AJ....138..332J} as reviewed in {\it Cosmicflows-2} \citep{2013AJ....146...86T}.  That earlier release contained 297 TRGB distances while here the number has grown to 384, a 29\% increase.

Currently roughly half of known galaxies brighter than $M_B=-12$ suspected to be within 10 Mpc have TRGB distances.  Interesting subsamples or individual targets have been discussed in separate publications \citep{2014ApJ...782....4K, 2014MNRAS.443.1281K, 2015ApJ...805..144K, 2015MNRAS.447L..85K, 2015ApJ...802L..25T}.  The methodology has been extended to the infrared, where the TRGB is bright and Galactic reddening is diminished \citep{2012ApJS..198....6D, 2014AJ....148....7W}.  

\begin{figure}[t]
\begin{center}
\includegraphics[scale=.4]{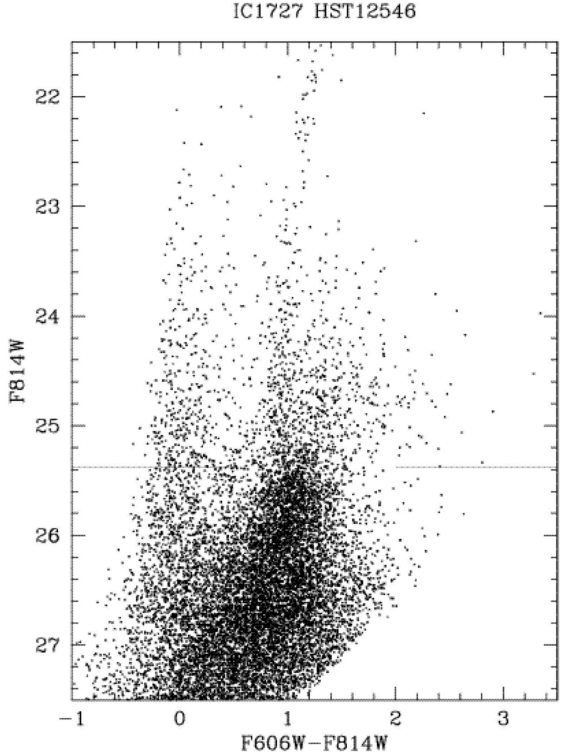}
\caption{Color-magnitude diagram for IC~1727.  The horizontal dashed line lies at the magnitude of the tip of the red giant branch and indicates a distance of 7.5 Mpc.}
\label{cmd}
\end{center}
\end{figure}

Figure~\ref{cmd} provides an example of a TRGB measurement with {\it Hubble Space Telescope} data.  In this case, there are prominent Pop I features in the main sequence at color F606W-F814W$\sim0$ and the red supergiant column at F814W$<25$ at F606W-F814W$\sim1$, and an abundance of intermediate age asymptotic giant branch stars above the TRGB.  The red giant branch itself is widened by metallicity and age mixing.  Nevertheless, the TRGB is defined at the level of 5\% in magnitude.  The fit is made with a maximum likelihood algorithm that incorporates the evaluation of superposed fake stars to determine completion and measurement uncertainties \citep{2006AJ....132.2729M}.  A weak metallicity$-$age dependent color term is applied  \citep{2007ApJ...661..815R}.  As a guide to the eye, the tick marks are at intervals of 0.2 on the magnitude scale, steps of 10\% in distance.

Knowledge of so many TRGB distances has been transformative.  Uncertainties in these distances are at the level of 5\%.  Group affiliations are unambiguous.  Meaningful measures can be made of individual peculiar velocities, $V_{pec} = V_{obs} - H_0 d$.  Here, $V_{obs}$ is the observed velocity and the cosmic expansion expectation velocity is the product of the Hubble Constant, $H_0$, and distance $d$.  At 10 Mpc an uncertainty of 5\% in distance transforms to an uncertainty in $V_{pec}$ of less than 40~\kms, considerably less than deviant motions of a few hundred \kms\ seen above and below the supergalactic plane and toward the Virgo Cluster \citep{2008ApJ...676..184T, 2014ApJ...782....4K, 2015ApJ...805..144K}.

\section{Type Ia Supernova Distances}
\label{sn}

Peculiar velocity uncertainties associated with a particular methodology grow linearly with redshift.  Uncertainties from Type Ia supernova (SNI$a$) measurements are at least a factor 2 less than those from the TF or FP estimates, so each one is four or more times as valuable.  Indeed, at velocities greater than 10,000~\kms\ to the 30,000~\kms\ limit of this catalog, SNI$a$ are making an increasingly dominant contribution.

{\it Cosmicflows-2} contained 309 SNI$a$ distances from five sources, with {\it Union-2} \citep{2010ApJ...716..712A} as backbone.   The most important new contribution is provided by \citet{2014ApJ...795...44R}.  Restricting to $z<0.1$, this new sample contains an overlap of 141 SNI$a$ with our earlier compilation and 58 new cases.  The substantial overlap provides a robust zero-point translation of Rest et al. distance moduli to the {\it Union-2} sample on the CF2 zero-point scale, $\mu_{union2}^{cf2} = \mu_{rest} - 0.167 $ ($\pm 0.007$).  The r.m.s. scatter between old and new measures is a satisfactory $\pm 0.079$, 4\% in distance. The zero point offset found for the 141 SNI$a$ in common with {\it Cosmicflows-2} is then applied to the 58 additional objects studied by Rest et al.

\cite{2015ApJS..219...13W} provide material for 29 additional distances to SNI$a$ within $z=0.1$.  The cases are all new so the zero-point match is less constrained.  
Walker et al. assert that their zero-point is set by the choice $H_0 = 70.8$~\kmsMpc, which is consistent with the distances we derive from the parameters they provide.  Our input zero-point is consistent with $H_0 = 74.4$~\kmsMpc\ \citep{2014ApJ...792..129N}.  Consequently, we decrease the Walker et al. distances by a factor 0.952, or 0.108 in the modulus.  

While no direct comparison can be made with an alternative distance estimate to a host galaxy for this sample,  a check of our proposed rescaling is afforded by events that have occurred in two groups with alternative distance estimates.  In the case of the NGC~5044 Group, nest 100019 in  \citet{2015AJ....149..171T}, one surface brightness fluctuation and three TF distances have a weighted average modulus of 32.56, in excellent agreement with the rescaled modulus of 32.63 for supernova $2013aiz$ in ESO~576-017.  However the situation is perplexing in the case of Abell~539 where the earlier supernova $2004ge$ \citep{2010AJ....139..120F} along with 10 TF and 25 FP (plus two measures in the current $Spitzer$ sample) are in good agreement at a modulus 35.29, with formal 95\% uncertainty $\pm 0.12$ (the modulus for $2004ge$ alone is 35.28).  The Walker et al. supernova $2011pn$, at modulus 35.68 is statistically consistent.  Supernova $2011ot$, at 36.12, is unexpectedly deviant.  While better agreement could be forced with an unseemly large adjustment to the Walker et al. scale, the substantial difference between the $2011ot$ and $2011pn$ moduli would remain.  It seems best to not place too much weight on individual measurements.

While the relative distances of the 29 systems in the Walker et al. sample are expected to be of similar high quality as other SNI$a$ measures, the zero-point linkage is subject to uncertainty at the level of $\sim 5\%$.  With the addition of material from the two new literature sources, we now have 391 SNI$a$ distances within $z=0.1$.  We need to check if the minor revisions of the TF calibration have slightly changed the zero point of the SNI$a$ scale, but before giving attention to that issue there is one more improvement to discuss.

\section{Group Catalog}
\label{nests}

Averaging over groups reduces uncertainties in both distances and velocities.  In the case of velocities, the averaging can include {\it all} known group members, not just the relatively few with measured distances.  The ensuing advantage is particularly evident in the case  of rich clusters with large velocity dispersions.

In the case of distances the interests are three-fold.  There is the clear advantage of reducing distance uncertainties.  Also, integrating over many groups, comparisons between methods constrain zero-point offsets.  And multiple measures within a group serve to isolate egregiously discordant distance estimates.

The group catalog used with the construction of {\it Cosmicflows-2} was poorly defined.  A new catalog \citep{2015AJ....149..171T} is more rigorous.  The input for the new group catalog is the 2MASS Redshift Survey (2MRS) sample of 43,526 galaxies brighter than $K_s=11.75$ \citep{2012ApJS..199...26H}.  This sample strongly overlaps the 6dFGS sample, the latter extending to a fainter 2MASS limit but only in the celestial south.  The 2MRS $K<11.75$ sample peaks in numbers by 10,000~\kms\ but extends to beyond the 6dFGS cutoff of 16,000~\kms.

Three-quarters of the galaxies in the present collection of distances associate with groups (called ``nests'') in the catalog of \citet{2015AJ....149..171T}.  The linkages are either direct galaxy matches or indirect fits to spatial and velocity criteria.  In the latter cases the criteria are specified by the total 2MASS $K$-band luminosities of the groups, corrected for missing light with distance.  The integrated group luminosities predict expectation velocity dispersions, $\sigma_p$, and virial dimensions characterized by the radius of second turnaround, $R_{2t}$.  Group linkages are contemplated if velocities are within $3\sigma_p$ of the group mean and projected locations are within $1.5 R_{2t}$.  As a second step, linkages are rejected if {\it both} velocities deviate by greater than $2\sigma_p$ of the mean and locations are outside $1.0 R_{2t}$.  Accordingly, 6348 of 8198 galaxies in the old {\it Cosmicflows-2} are linked to nests (77\%), 6157 of 8885 galaxies in 6dFGS are linked (69\%), 1780 of 2281 galaxies in the $Spitzer$ sample are linked (78\%), and 293 of 391 SNI$a$ hosts are linked (75\%).

In order to ferret out bad data, we first look at consistency in distances between direct matches of alternate sources.  There are 1524 galaxies with TF distances that are alternatively based on $I$-band photometry as reported in {\it Cosmicflows-2} or $Spitzer$ [3.6] photometry introduced here.  There were 17 strongly deviant cases (moduli differences $>0.8$ mag). Nine of these are nearby galaxies with TRGB, Cepheid, SNI$a$, or SBF distances.  Most of these galaxies are low luminosity, rather irregular systems poorly suited for the TF methodology.  The nine $Spitzer$ measures are rejected in favor of the other available distances.

In the other 8 cases the discordances are between the new $Spitzer$ sample and the earlier $I$-band material.  Aside from the sources of photometry, the only substantial differences are the assumed inclinations (recall Section 2.1).  In all 8 cases, it is obvious from inspection of images that the new inclinations are more representative than the values used previously.  The $Spitzer$ distances are retained and the distances in {\it Cosmicflows-2} for these 8 galaxies are discarded.  In the 1507 remaining matches, $<\mu_{spit}-\mu_{cf2}>=-0.035\pm0.004$ with r.m.s. scatter of only $\pm 0.167$ since usually only the photometry changes, and occasionally the inclinations.

There is a substantial incidence of SNI$a$ in groups and multiple occurrences in the same nest afford a test of the dispersion
in supernova distance measures.  There are three SNI$a$ in five nests and two such events in six nests in our sample.  The estimated distance moduli for the 27 events can be compared jointly by calculating the r.m.s. dispersion from the mean distance moduli of their nests of residence.  It is found that $<\mu_{nest_j}^{sn_i} - \bar{\mu}_{nest_j}> = 0.02 \pm 0.21$ (10\% scatter in distance) for 27 events.  There is one strongly deviant case: sn2011$ot$ in Abell 539, already noted in Section~\ref{sn}, differs  from the group mean by  0.83 mag ($4 \sigma$).  The three supernovae in the cluster have a dispersion in modulus of 0.42 mag.  Two supernovae ostensibly in Abell 1367, sn2006$bd$ and sn2007$ci$, differ in distance modulus by 0.91, 52\% in distance.

\begin{figure}[t]
\begin{center}
\includegraphics[scale=.4]{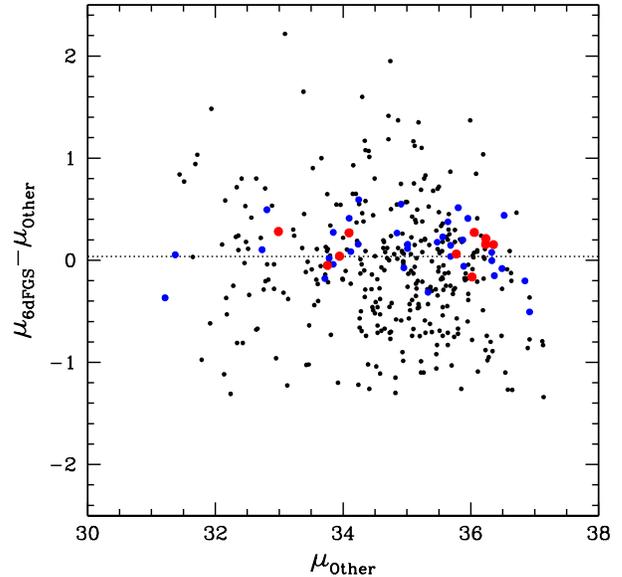}
\caption{Comparison of 6dFGS type adjusted distance moduli with alternative measures.  The 10 points in red represent nests with at least 10 distance measures by both 6dFGS and alternatives. The additional 34 points in blue represent nests with $4-9$ distance measures by each. For cases represented by black points there are three or less measures by at least one of 6dFGS or other.}
\label{deldm}
\end{center}
\end{figure}

Figure~\ref{deldm} revisits the issue of the integration of the zero point for the 6dFGS FP distances with the other material.  A comparison is made of distances to 381 \citet{2015AJ....149..171T} nests.  Averaging across all nests, with appropriate weights, there is a mean difference $<\mu_{6df}-\mu_{other}>=\Delta \mu=0.036 \pm 0.020$.  

\subsection{The Virgo Cluster}
\label{virgo}

An important rung in the extragalactic distance ladder, the Virgo Cluster, presents an interesting example of insidious projection problems.  There is always cause to worry about group contamination, especially at low redshift where galaxies with similar velocities can be at different distances that are large in a relative sense.  The Virgo Cluster is close, indeed, close enough that with the most accurate distance methodologies $-$ Cepheids, TRGB, SNI$a$, SBF $-$ foreground and background objects can be clearly distinguished.  By misfortune there is a prominent Fornax-scale cluster, called Virgo W by \citet{1961ApJS....6..213D}, that lies at roughly twice the Virgo distance, positioned in projection such that the virial edges of the two clusters abut.  The mean velocity of Virgo W is larger than for Virgo but the velocities within the dispersions of the two clusters completely overlap.  Membership assignments with either of the two entities require good distances.

Although the main part of Virgo W projects outside the virial radius of the Virgo Cluster, there are known Virgo interlopers.  The M Cloud \citep{1984ApJ...282...19F} lies at the background distance of Virgo W and projects fully onto the Virgo Cluster.  Then there is the less substantial contaminant Virgo W$^{\prime}$ \citep{1961ApJS....6..213D} that lies 50\% farther than Virgo and is fully within the Virgo spatial and velocity window.

This situation is discussed at greater length in the Virgo section on the paper on galaxy groups by \citet{2015AJ....149...54T}. The point was made that most of the known contamination lies within an area representing $\sim 40\%$ of the surface of Virgo.  This perception had already informed the TF calibration analysis of \citet{2000ApJ...533..744T} and of subsequent studies \citep{2012ApJ...749...78T, 2013ApJ...765...94S, 2014MNRAS.444..527S, 2014ApJ...792..129N}.  Only galaxies lying in the $\sim 60\%$ of the cluster without much contamination were considered as TF calibrators.  The same strategy was followed in this new study.  There is the consequence that the new collection of distances provides information on many more galaxies in and around Virgo than are used in the calibration. It is of interest to examine  what they tell us.

\begin{figure}[!]
\begin{center}
\includegraphics[scale=.4]{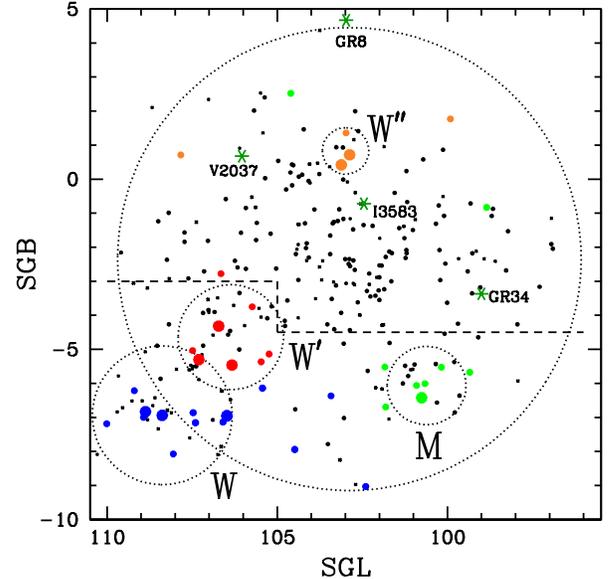}
\caption{Known interloper galaxies with distances that distinguish them from Virgo Cluster members.  Blue and green symbols identify galaxies in or related to the Virgo W Cluster and M Cloud at twice the distance of the Virgo Cluster.  Red and orange symbols identify galaxies associated with the Virgo W$^{\prime}$ Group midway between Virgo and Virgo W.  Large symbols identify systems with distance measures more accurate than 10\% while small symbols identify systems with distance measures with $\sim 20\%$ accuracy.  The dark green stars locate dwarf galaxies to the foreground of the cluster.  The black dots are at the positions of galaxies with 2MASS K magnitudes brighter than 11.75 lacking distance information.  The second turnaround radii for the separate groups are indicated by dotted circles.  The Virgo sample for the TF template is drawn exclusively from above the dashed lines.}
\label{notvir}
\end{center}
\end{figure}

Figure~\ref{notvir} uses color to identify the projected locations of interloper galaxies imposed onto the Virgo Cluster.  The circle with $6.8^{\circ}$ radius is the projected second turnaround radius \citep{2015AJ....149...54T}, a proxy for the virial radius.  The main features in projection, Virgo W and W$^{\prime}$ and the M Cloud are identified with their second turnaround domains.  Distances to the galaxies in colors and large symbols come from Cepheid measurements \citep{2001ApJ...553...47F} or from SBF observations \citep{2001ApJ...546..681T, 2007ApJ...655..144M, 2009ApJ...694..556B}.  These galaxies lie either with Virgo W at twice the Virgo distance or half way between Virgo and Virgo W with Virgo W$^{\prime}$.  See the figure caption for details.  So far, three foreground objects at $\sim 9$~Mpc have turned up from HST TRGB observations \citep{2014ApJ...782....4K}.  The dotted lines in the figure show the boundary isolating the domain of the Virgo TF calibrators $-$ they all lie at positive SGB with respect to the boundary.

\begin{figure}[t]
\begin{center}
\includegraphics[scale=.4]{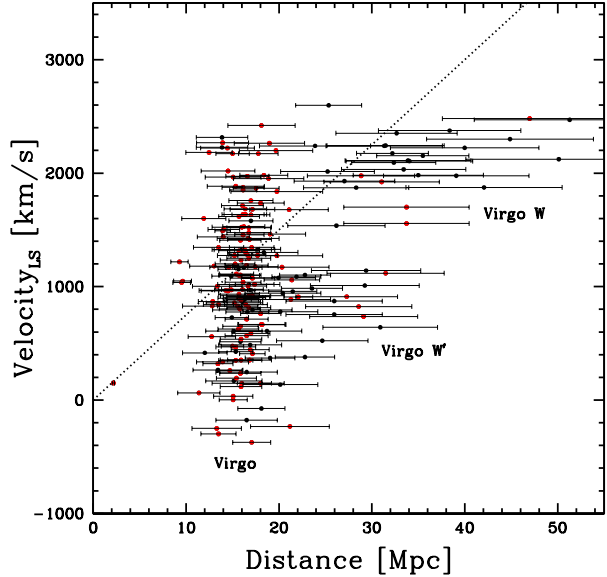}
\caption{Distance vs. velocity for galaxies within a radius of $8^{\circ}$ of the Virgo Cluster with distance estimates.  Red/black: those with supergalactic latitude values above/below the dashed boundary in Fig.~\ref{notvir}.  Dotted line is Hubble law with $H_0=75$.}
\label{virdv}
\end{center}
\end{figure}

The uncertainties in distances from the TF and FP correlations do not allow for a clean rejection of contaminants.  However, from Figure~\ref{virdv} the proposition of confusion is evident and, to a reasonable degree can be disentangled.  Galaxies with local sheet velocities below 400~\kms\ are almost inevitably in the cluster, save for a few foreground dwarfs.  Galaxies related to the background Virgo W have velocities greater than 1500~\kms\ and at 1.5 mag displacement are separable with distance information in all but pathological cases.  The galaxies in and around Virgo W$^{\prime}$ are more problematic, with velocities near the Virgo mean and distance displacements of less than $2 \sigma$ for FP and TF measures.  However the Virgo W$^{\prime}$ feature is not populous.

\begin{figure}[!]
\begin{center}
\includegraphics[scale=.4]{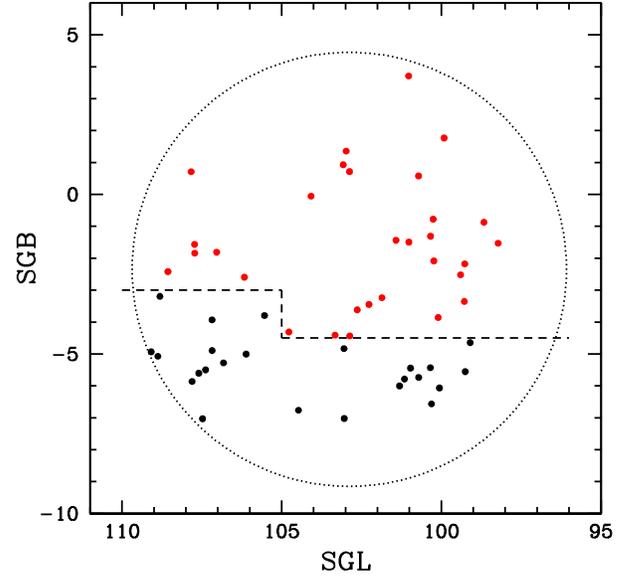}
\includegraphics[scale=.4]{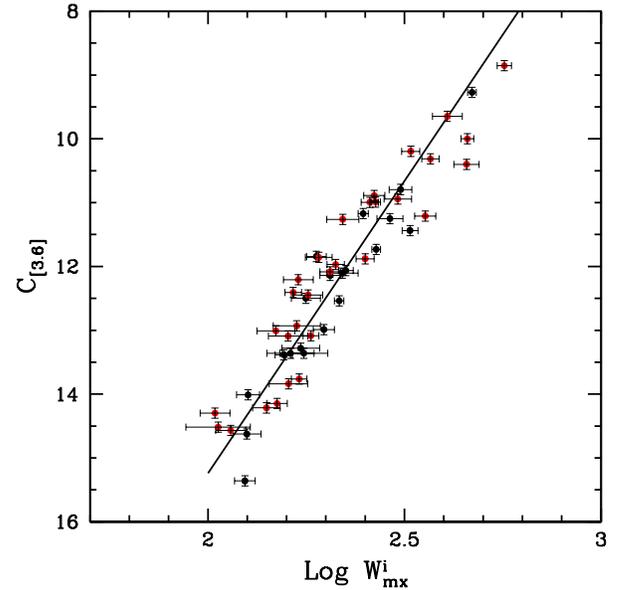}
\caption{Galaxies associated with the Virgo Cluster with TF distances.  Those in red, all above the dashed line in the top panel in the region relatively free of interloper contamination, contribute to the TF template calibration.  Those in black are not used in the calibration.  The combined samples contribute to the TF plot in the lower panel.  The solid line has the 13 cluster template slope.}
\label{virtf}
\end{center}
\end{figure}

It remains to compare distance results for the Virgo calibrator sample and the ensemble attributed to the cluster.  The targets are seen in Figure~\ref{virtf}.  The calibrators are in red and others with distances from the TF correlation attributed to the cluster, not the background, are in black.  The mean modulus for the calibrator sample of 30 galaxies is $30.96 \pm 0.11$ with rms scatter 0.58.  The mean modulus for the entire sample of 50 galaxies within the second turnaround radius is $31.04 \pm 0.08$ with rms scatter 0.53.  Including 161 distance measures by all techniques that we use for the Virgo Cluster proper, the modulus is 31.01 (15.9 Mpc), with rms scatter 0.27 and nominal uncertainty 0.02 mag.

\section{Homogenized Zero Point and $H_0$}

Our distance scale lattice involves three regimes.  Nearby, within $\sim20$~Mpc, our construction depends on Cepheid and TRGB distances, based on Pop I and Pop II foundations respectively and seen to agree. It is assumed for the Pop I scale that the Large Magellanic Cloud has a modulus 18.48  \citep{2012ApJ...758...24F}. The Pop II scale is set by Horizontal Branch fits to M33, NGC 185, IC 1613, and Sculptor and Fornax dwarfs as described by   \citet{2007ApJ...661..815R}.  Distances by both methods agree with the maser distance to NGC 4258 \citep{2013ApJ...775...13H}.

The SBF scale is tied directly to the Cepheid scale \citep{2001ApJ...546..681T, 2009ApJ...694..556B, 2010ApJ...724..657B} and confirmed on average to be compatible with our other distances.

The TF and FP methods are operative in the broad regime $20-200$~Mpc.  The TF calibration is built through the 13 cluster slope template and the Cepheid-TRGB zero point constraints.  There is a small difference in zero point between the scale established by the $Spitzer$ sample and the earlier CF2 sample.  The difference is evaluated in two ways.  One way is to determine the difference in distance moduli $\Delta \mu = \mu_{spit} - \mu_{cf2}$ averaged over the 13 template calibrators.  We find $\Delta \mu =-0.016$ with r.m.s. scatter $\pm 0.099$.  Alternatively, the difference in distance moduli can be averaged over all 1507 galaxies in common, whence $\Delta \mu = -0.035$ with scatter $\pm 0.167$.  These offsets agree to within a half standard deviation.  We accept a straight average of the two values, $\Delta \mu = -0.025$.  The $Spitzer$ recalibration results in distances $1.2\%$ smaller on average.

The bulk of FP material in the CF2 compilation was restricted to clusters and the linkage of that sample was provided by cluster comparisons \citep{2013AJ....146...86T}.  The 6dGS FP distances were linked to the TF scale through clusters and individual galaxy matches. However, since the overlap of 6dFGS is limited, their contributions are excluded from the ladder to define $H_0$.

SNI$a$, though sparse in coverage, are seen usefully at all relevant distances.  The 309 SNI$a$ available for {\it Cosmicflows-2} were given a compatible zero point.  Now, as described in Section~\ref{sn}, 58 additional SNI$a$ are added from \citet{2014ApJ...795...44R} with a statistically robust fit to the same zero point and 29 SNI$a$ are added from \citet{2015ApJS..219...13W} with a less secure zero point fit.  What is now required is a fit of the ensemble of SNI$a$ to the new slightly revised calibration.

The comparison is enabled by the nests.  As described in Section~\ref{nests}, 75\% of SNI$a$ hosts can be linked to the 2MASS nests of \citet{2015AJ....149..171T}.  We then look for alternative distances for these nests within the $Spitzer$ TF sample or the CF2 sample, excluding the SNI$a$ contributions.  With the large nests there can be multiple distance estimates by alternative methods and even occasions with multiple SNI$a$; there are five nests with three SNI$a$ and six with two.  Multiple measures can be averaged.  Then in addition, there are six instances of SNI$a$ hosts, not in established nests, but individually with an alternative distance estimate.

\begin{figure}[!]
\begin{center}
\includegraphics[scale=.4]{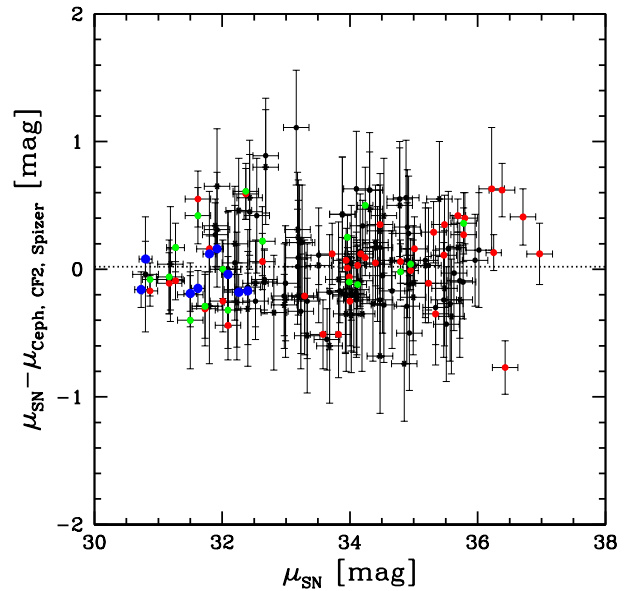}
\caption{Fine-tuning of the SNI$a$ zero point.  SNI$a$ distance moduli are given on the abscissa with the zero point established for the ensemble of SNI$a$ in CF2.  Differences between the SNI$a$ distance moduli and distance moduli given by alternate methods are given on the ordinate axis on the CF2 scale.  Large blue symbols: individual Cepheid$-$SNI$a$ comparisons.  Black and green symbols: comparisons involving 1 or 2 alternative distance measures against a SNI$a$ measure in the same nest; black if from CF2 and green if from the new $Spitzer$ compilation.  Red symbols: comparisons involving 3 or more alternative measures (either from CF2 or Spitzer) against 1 to 3 SNI$a$ measures within the same nest.  The weighted average involving 9 individual Cepheid matches, 102 nests with CF2 measures, and 66 nests with $Spitzer$ measures, implies SNI$a$ moduli on the CF2 scale are high by 0.021 mag.}
\label{sncal}
\end{center}
\end{figure}

Comparisons are shown in Figure~\ref{sncal}.  SNI$a$ moduli, on the CF2 scale, are plotted against the difference, $\mu_{sn} - \mu_{other}$, where $\mu_{other}$ can alternatively be from the $Spitzer$ TF sample or the re-calibrated CF2 sample.  The comparisons are in red if there are three or more alternative distances (over 100 in each of Virgo and Coma clusters and greater than 10 in 18 other cases), with the vertical error bars indicative of the number and quality of measures.  Cases with one or two measures are black if from CF2 or green if from Spitzer.  It is to be appreciated that the $Spitzer$ and CF2 samples are not fully independent, since many of the SNI$a$, nests, and individual galaxies involved are the same.

Finally for the comparisons, the large blue symbols identify 9 cases where the SNI$a$ host has a Cepheid distance.  These cases plus the nest matches are averaged, taking due account of weights and the quasi-dependencies between $Spitzer$ and CF2 samples, leading to the determination $\mu_{sn} - \mu_{other} = 0.021 \pm 0.023$, a statistical increase in distances of $1.0\%$.  The offset from the earlier CF2 scale is only $1\sigma$, but an adjustment of this observed amplitude is applied to the SNI$a$ scale.

It must be noted that a fit solely to the 9 hosts of SNI$a$ events with Cepheid distances gives $\mu_{sn} - \mu_{cepheid} = -0.059 \pm 0.047$.  Accepting this scale would increase the SNI$a$ scale by 0.080 in the modulus, an effect that would decrease $H_0$ by 3.8\%.  The 9 SNI$a$$-$Cepheid matches considered here strongly overlaps with the \citet{2011ApJ...730..119R} sample (we remove sn$1981B$ and add sn$1999by$ and sn$2006X$).  It is seen in Fig.~\ref{sncal} that the direct SNI$a$$-$Cepheid comparisons are offset from the mean but not by a statistically significant amount. The Riess et al. article that focuses on a SNI$a$ calibration with Cepheids alone is titled "A 3\% Solution".  Precision cosmology is a fraught quest.   

Returning to the \citet{2014ApJ...795...44R} compilation, this extensive sample extends beyond $z=0.1$ to mid and high redshifts and, hence, can provide a bridge to the global Hubble Constant with minimal distortion from peculiar velocities.  The component of the Rest et al. compilation of interest is a new sample of SNI$a$ based on cadenced photometry with {\it Pan-STARRS}  \citep{2010SPIE.7733E..0EK}. The Hubble Constant that is obtained with this new sample complements the parallel determination using the {\it Union-2} catalog of SNI$a$ distances \citep{2010ApJ...716..712A} that is the foundation for peculiar velocity estimates in {\it Cosmicflows-2} \citep{2012ApJ...758L..12S, 2014ApJ...792..129N}.  

\begin{figure}[t]
\begin{center} 
\includegraphics[scale=.4]{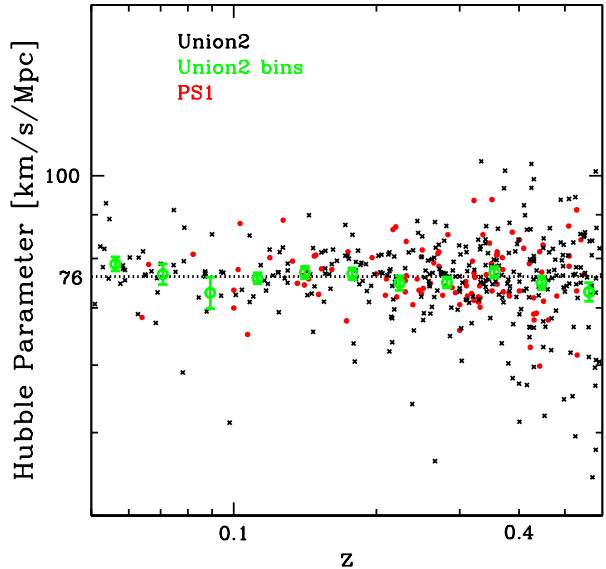}
\caption{Fit to individual values of the Hubble parameters for supernovae in the {\it Union-2} (black crosses) and PS1(red circles) samples in the interval $0.05 < z < 0.6$.  Binned {\it Union-2} values are shown as open green circles. The logarithmically averaged value of $H_0=76.2$~\kmsMpc\ is shown by the dashed line.}
\label{ps1}
\end{center}
\end{figure}

Figure~\ref{ps1} shows the dependence of the Hubble parameter, $H_i$, as a function of redshift for the two alternative SNI$a$ samples, {\it Union-2} and {\it Pan-STARRS} (PS1).  The parameter $H_i$ has the cosmological corrections of Eq.~\ref{hi}, assuming $\Omega_m = 0.27$ and a flat cosmic topology.  Separately, the {\it Union-2} sample is consistent with $H_0 = 75.9$~\kmsMpc\ and the PS1 sample implies $H_0 = 76.4$~\kmsMpc.  We take an average and accept $H_0 = 76.2$~\kmsMpc.  There is a mild dependence on the cosmological density parameter, with $H_0$ increased/decreased one unit if $\Omega_m$ is decreased/increased 0.04.

There are both random and systematic uncertainties in the estimate of $H_0$.  The largest random uncertainty ($\pm 0.079$ in the modulus) lies with the zero point calibration to the TF relation based on 33 galaxies with Cepheid or TRGB distances.  The SNI$a$ zero point is directly established through 9 Cepheid distances and less directly established through over 100 group affiliations and has an uncertainty of $\pm 0.046$.  Then there is an uncertainty of $\pm 0.030$ for the fit to the mean Hubble parameter in Figure~\ref{ps1}.  Added in quadrature, the random error is $\pm 0.096$, 4.5\% in distance and affect on $H_0$.  Systematic effects are harder to quantify, but the components of concern and estimates of uncertainties are the Cepheid and TRGB zero points ($\pm 0.05$), possible variations in SNI$a$ properties that manifest in distance \citep{2015ApJ...802...20R} ($\pm 0.05$), and uncertainties in the cosmological model corrections ($\pm 0.027$).  Systematics crudely add in quadrature to 0.076, 3.5\% in distance and $H_0$.

\section{The Catalog}

Information is gathered on two levels.  At level one are individual galaxies.  In cases with multiple distance measurements, averaged moduli are determined with weights appropriate to the different inputs.  At level two are groups, specifically the nests of \citet{2015AJ....149..171T} which supplies group averaged positions and velocities plus summed $K$ band luminosities and estimated masses.  Group distance moduli are derived by weighted averaging of all the moduli available for group members.

The group catalog \citep{2015AJ....149..171T} was built exclusively from 2MRS, the 2MASS $K<11.75$ redshift sample \citep{2012ApJS..199...26H}.  The catalog includes "nests"  as small as only one member.  {\it Cosmicflows-3} contains entries that are not in 2MRS.  It was described in Section~\ref{nests} that roughly three-quarters of {\it Cosmicflows-3} galaxies are assigned to a 2MRS nest.  Galaxies that define the distance to a nest and galaxies that define the other properties (position, velocity, luminosity) need not be identical.  Then additionally, there are galaxies with distances that have no nest assignment.  These cases are taken to be singles.

The catalog is presented as two tables, abbreviated in this text but available on-line in complete form.  Table~\ref{grouped} gives summary group information for 11,508 nests, 1704 with two or more distances and 9804 singles.  The Virgo Cluster (nest 100002) has the most galaxies with measured distances with 161.  There are 125 groups with 10 or more galaxies with distances.  Table~\ref{individual} couples the individual and group information.  In this table, a row is dedicated to each of the 17,669 galaxies with distance measurements.  The first 41 columns provide information on the specific galaxy, including distances from alternative sources.  Then the last 30 columns give information on the associated nest, drawn from Table~\ref{grouped}.  The material in these latter columns is the same for every member of a given nest. 

The distance moduli, both for individual galaxies and again for groups, are weighted averages with individual weights $w_i = 1/\epsilon_i^2$ where the uncertainty in a modulus is $\epsilon_i$, giving total weights $w_t = \Sigma_i^N w_i$, and error on the averaged modulus $\epsilon_{\mu} = 1/w_t^{1/2}$.  For groups with many distance measures this formal error is as small as 0.02 mag, ~3 times smaller than reasonable expectations of systematic errors.

\noindent
{\it Description of Tables~\ref{grouped} and \ref{individual}.}  In the case of the group catalog, Table~\ref{grouped}, columns $1-30$ are mimicked in columns $42-71$ of Table~\ref{individual}, the catalog of individual sources.  The following is a description of the columns in Table~\ref{individual}.  This table is also available, with updates, at the {\it Extragalactic Distance Database}.\footnote{http://edd.ifa.hawaii.edu}

\begin{enumerate}
\item  PGC: Principal Galaxies Catalog ID \citep{2014A&A...570A..13M}

\item  $d$: Luminosity distance measurement for individual galaxy.  Weighted average of the distance moduli if more than one source. [Mpc]

\item  $N_d$: Numbers of sources of distance measures for the individual galaxy.

\item  $<\mu>$: Luminosity distance modulus measurement for individual galaxy.  Weighted average if more than one source. [mag]

\item  $\epsilon_{\mu}$: One standard deviation uncertainty in distance modulus.  Caution: Different from CF2 entry where the fractional error in distance is quoted. [mag]

\item  C: indicates the availability of a Cepheid measurement.

\item  T: indicates the availability of a Tip of the Red Giant Branch measurement from an HST observation and reduced in a standard way \citep{2009AJ....138..332J}.

\item  L: indicates the availability of a Tip of the Red Giant Branch measurement extracted from the literature.

\item  M: indicates the availability of a high quality miscellaneous measurement (RR Lyrae, Horizontal Branch, Eclipsing Binary, Maser).

\item  S: indicates the availability of a Surface Brightness Fluctuation measurement.

\item  N: indicates the availability of a Type Ia Supernova measurement in the host galaxy.

\item  H: indicates the availability of a TF relation measurement based on optical photometry, as reported in CF2.

\item  I: indicates the availability of a TF relation measurement based on photometry at 3.6~$\mu$m with {\it Spitzer Space Telescope}.

\item  F: indicates the availability of a Fundamental Plane measurement from the ENEAR, EFAR, or SMAC experiments as reported in CF2.

\item  P: indicates a Fundamental Plane measurement from 6dFGS \citep{2014MNRAS.445.2677S}.

\item  $\mu_{cf2}$: Distance modulus carried over from {\it Cosmicflows-2}. [mag]

\item  $\epsilon_{\mu}^{cf2}$: One standard deviation uncertainty in distance modulus carried over from CF2. [mag]

\item  SN: Supernova identification.

\item  $N_{sn}$: Number of separate analyses of the supernova.

\item  $\mu_{sn}$: Luminosity distance modulus of supernova averaged over all contributions. [mag]

\item  $\mu_{spit}$: Luminosity distance modulus determined from the TF correlation using $Spitzer$ [3.6] band photometry. [mag]

\item  $\epsilon_{\mu}^{spit}$: One standard deviation uncertainty in distance modulus determined from TF relation using $Spitzer$ [3.6] photometry.  Uncertainty 0.45 if from color corrected relation; 0.54 if color information lacking. [mag]

\item  $\mu_{6df}$: Luminosity distance modulus from 6dFGS  \citep{2014MNRAS.445.2677S} Fundamental Plane with {\it Cosmicflows-3} zero point. [mag]

\item  $\epsilon_{\mu}^{6df}$: One standard deviation uncertainty in distance modulus determined from 6dFGS Fundamental Plane. [mag]

\item $M_t$: 6dFGS morphology type code (Campbell+ 2015).

\item  $RAJ$: Right ascension, epoch 2000. [hhmmss.s]

\item  $DecJ$: Declination, epoch 2000. [ddmmss]

\item  $Glon$: Galactic longitude. [deg]

\item  $Glat$: Galactic latitude. [deg]

\item  $SGL$: Supergalactic longitude. [deg]

\item  $SGB$: Supergalactic latitude. [deg]

\item  $Ty$: Morphological type; RC3 numeric code \citep{2014A&A...570A..13M}.

\item  $A_{sf}$: Reddening at B band \citep{2011ApJ...737..103S}. [mag]

\item  $B_t$: Total $B$ magnitude from LEDA. [mag]

\item  $K_s$: 2MASS $K_s$ magnitude with corrections from \citet{2011MNRAS.416.2840L}. [mag]

\item  $V_h$: Heliocentric velocity. [\kms]

\item  $V_{gsr}$: Velocity in Galactic standard of rest; circular velocity at Sun of 239 \kms; total velocity 251 \kms\ toward $\ell=90$,  $b=0$ \citep{2012ApJ...753....8V}. [\kms]

\item  $V_{LS}$: Velocity in Local Sheet standard of rest \citep{2008ApJ...676..184T}. [\kms]

\item  $V_{cmb}$: Velocity in CMB standard of rest \citep{1996ApJ...473..576F}. [\kms]

\item  $V_{mod}$: Velocity in CMB standard of rest adjusted in accordance with a cosmological model with $\Omega_{matter}=0.27$ and $\Omega_{\Lambda}=0.73$. [\kms]

\item  Name: Common name.

\item  Nest: 2MASS "nest" group identification \citep{2015AJ....149..171T}.

\item  $N_d^{gp}$: Number of galaxies in group with distance measures.

\item  $<\mu>^{gp}$: Luminosity distance modulus to group; weighted over all contributions. [mag]

\item  $\epsilon_{\mu}^{gp}$: One standard deviation uncertainty in group distance modulus. [mag]

\item  $d^{gp}$: Luminosity distance to group; weighted average of distance moduli. [Mpc]

\item  Abell: Abell Cluster identification; ASxxx are from the southern supplement list.

\item  Group Name: Alternative name for group or cluster.

\item  $N_v$: Number of galaxies in group with positions and velocities in the 2MRS catalog \citep{2012ApJS..199...26H}.

\item  PGC1: Principal Galaxies Catalog ID of brightest group member.

\item  $Glon^{gp}$: Galactic longitude of group. [deg]

\item  $Glat^{gp}$: Galactic latitude of group. [deg]

\item  $SGL^{gp}$:  Supergalactic longitude of group; luminosity weighted (2MASS $K_s$). [deg]

\item  $SGB^{gp}$:  Supergalactic latitude of group; luminosity weighted (2MASS $K_s$). [deg]

\item  $Log L^{gp}$: Log summed $K_s$ luminosity of group; adjusted by correction factor for lost light \citep{2015AJ....149..171T}.  Assumed distance given by group velocity $V_{mod}^{gp}$ and $H_0=75$. [log $L^K_{\odot}$]

\item  $cf$: Luminosity selection function correction factor.

\item  $\sigma_p$: Projected velocity dispersion anticipated by corrected intrinsic luminosity \citep{2015AJ....149...54T}. [\kms]

\item  $R_{2t}$: Projected second turnaround radius anticipated by corrected intrinsic luminosity \citep{2015AJ....149...54T}.  Assumed distance given by group velocity $V_{mod}^{gp}$ and $H_0=75$. [Mpc]

\item  $V_h^{gp}$: Group heliocentric velocity. [\kms]

\item  $V_{gsr}^{gp}$: Group velocity in Galactic standard of rest; circular velocity at Sun of 239 \kms; total velocity 251 \kms\ toward $\ell=90$,  $b=0$ \citep{2012ApJ...753....8V}. [\kms]

\item  $V_{LS}^{gp}$: Group velocity in Local Sheet standard of rest \citep{2008ApJ...676..184T}. [\kms]

\item  $V_{cmb}^{gp}$: Group velocity in CMB standard of rest \citep{1996ApJ...473..576F}. [\kms]

\item  $V_{mod}^{gp}$: Group velocity in CMB standard of rest adjusted in accordance with a cosmological model with $\Omega_{matter}=0.27$ and $\Omega_{\Lambda}=0.73$. [\kms]

\item  $V_{rms}$: Group velocity dispersion. [\kms]

\item  $M_{12}^{bw}$: Group mass in units of $10^{12}~\Msun$ from virial theorem with bi-weight dispersion and radius parameters \citep{2015AJ....149..171T}.  Assumed distance given by group velocity $V_{mod}^{gp}$ and $H_0=75$. [$10^{12}~\Msun$]

\item  $M_{12}^{L}$: Group mass in units of $10^{12}~\Msun$ based on corrected luminosity and $M/L$ prescription \citep{2015AJ....149..171T}.  Assumed distance given by group velocity $V_{mod}^{gp}$ and $H_0=75$. [$10^{12}~\Msun$]

\item  LDC: Low density group ID \citep{2007ApJ...655..790C}.

\item  HDC: High density group ID \citep{2007ApJ...655..790C}.

\item  2M++: Group ID from 2MASS++ catalog \citep{2011MNRAS.416.2840L}.

\item  M\&K: Group ID from catalog by \citep{2011MNRAS.412.2498M}.

\item  Icnt: Internal reference ID.

\end{enumerate}

\section{Discussion}

\begin{figure*}[!]
\begin{center} 
\includegraphics[scale=.85]{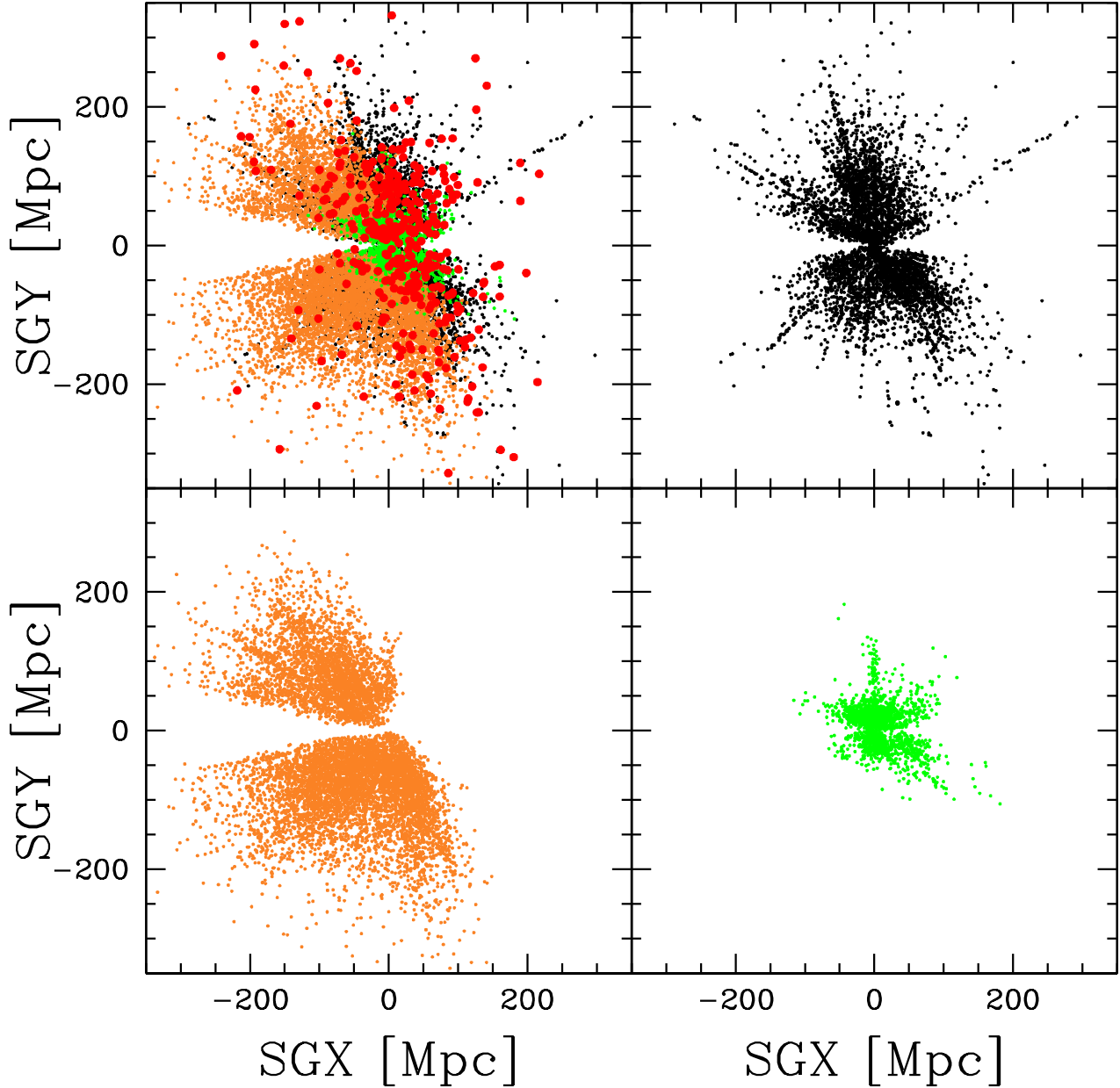}
\caption{Projection of individual CF3 distances into supergalactic SGX vs. SGY.  Upper left: all galaxies; the earlier CF2 sample is represented in black, the 6dFGS contribution is orange, the $Spitzer$ TF sample is green, and SNI$a$ hosts are shown in red.  Upper right: only the earlier CF2 sample.  Lower left: only the 6dFGS sample. Lower right: only the $Spitzer$ TF sample.}
\label{xy4}
\end{center}
\end{figure*}

\begin{figure*}[!]
\begin{center} 
\includegraphics[scale=.85]{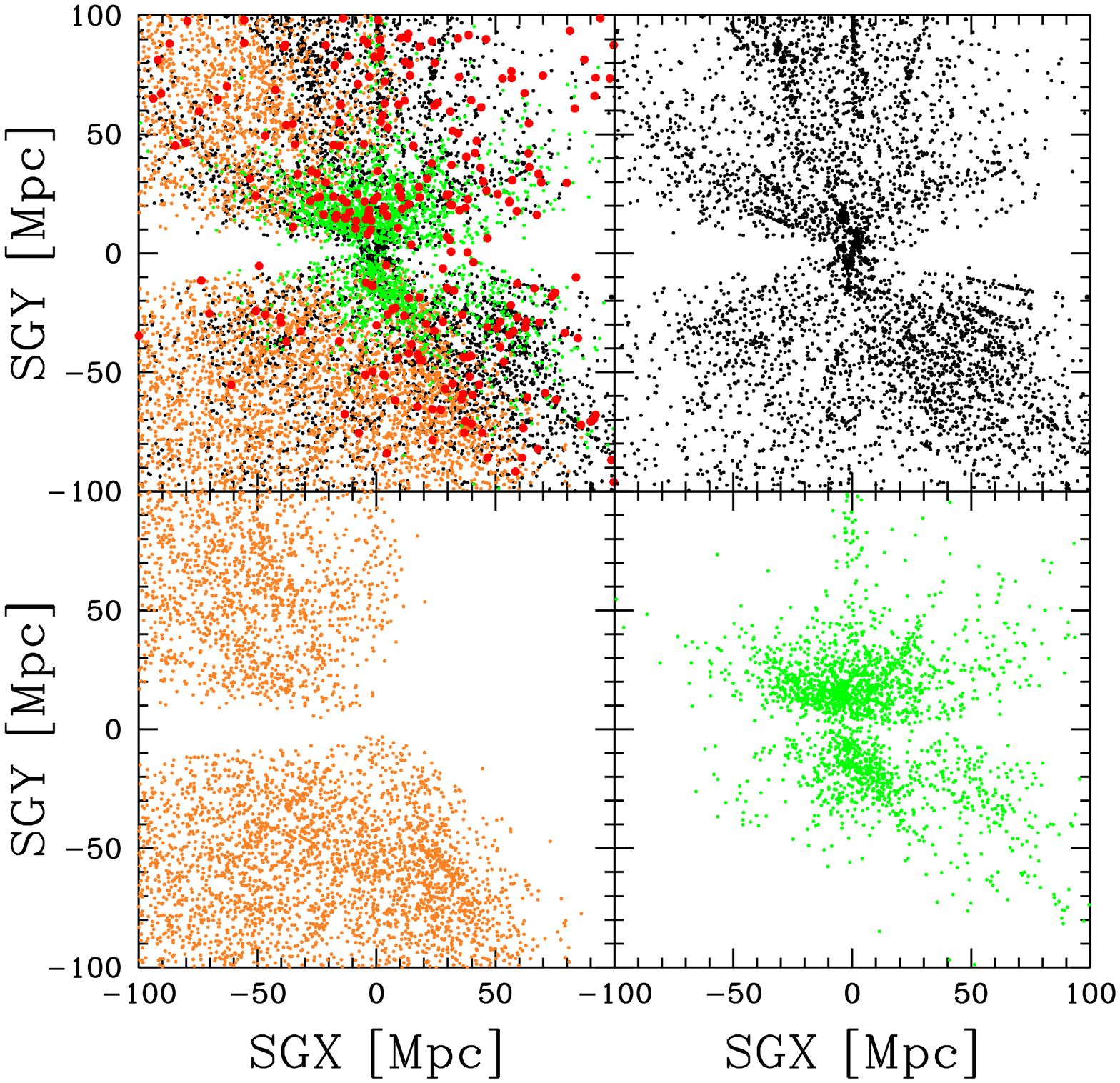}
\caption{Zoom of Fig.~\ref{xy4}.}
\label{xy4zoom100}
\end{center}
\end{figure*}

\begin{figure}[!]
\begin{center} 
\includegraphics[scale=.4]{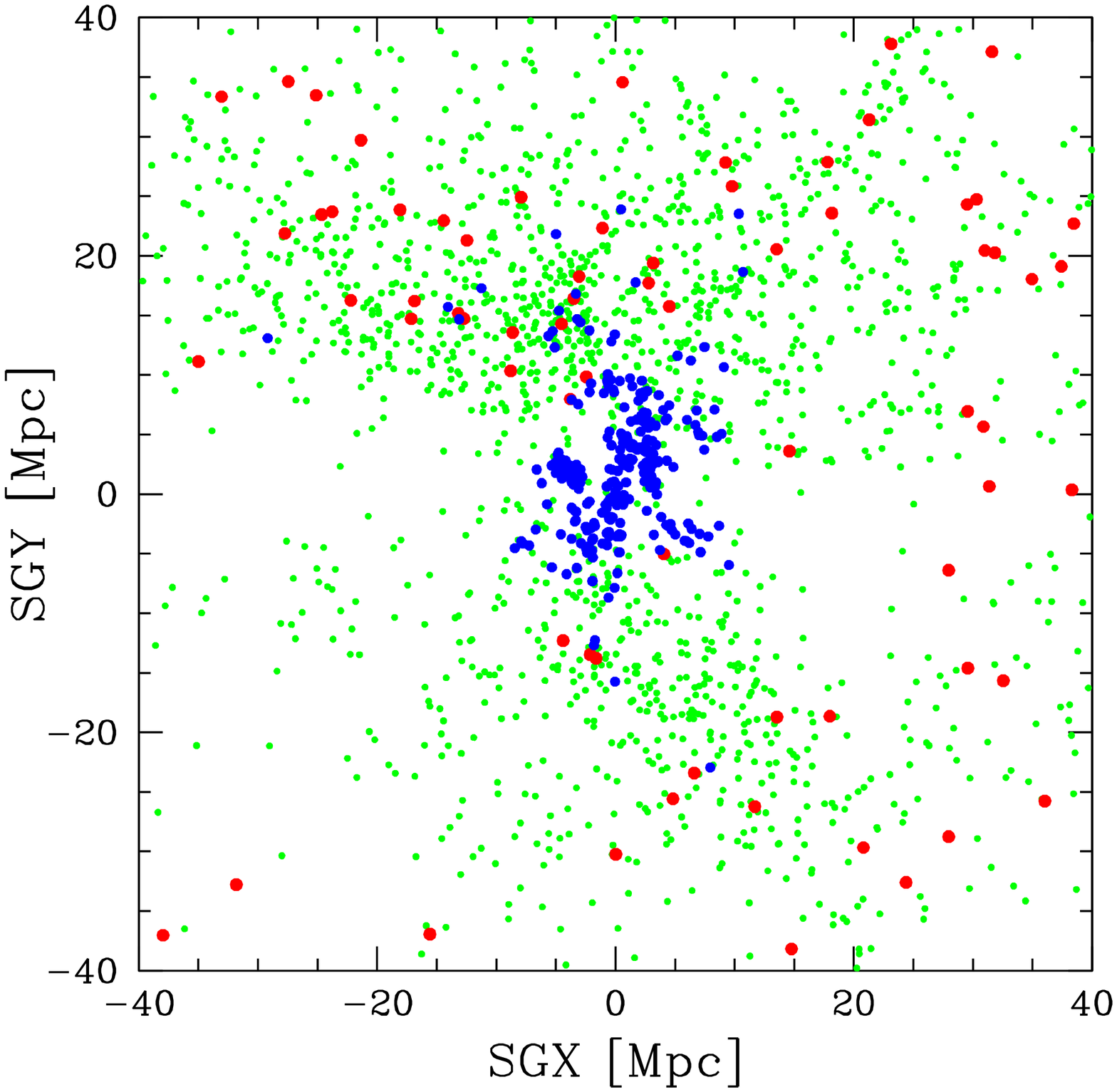}
\caption{Further zoom of lower right panel of Figs. \ref{xy4} and \ref{xy4zoom100} with the inclusion of SNI$a$ hosts in red and contributions from Cepheid and TRGB measurements in blue.}
\label{xyzoom40}
\end{center}
\end{figure}

The sky coverage is heterogeneous as evident in Figures \ref{xy4} $-$ \ref{xyzoom40}, which show  the projected distribution of the 17,669 galaxies in {\it Cosmicflows-3} on three different scales.  Figure~\ref{xy4} captures essentially the full domain of the catalog, in the ensemble in the upper left panel and by components in the other panels.  The numerical improvement of CF3 over CF2 is directly seen by the difference between the top left and right panels.  The two major new components are shown separately in the bottom panels.  Figure~\ref{xy4zoom100} is a zoom to the inner regions of these same panels.  Figure~\ref{xyzoom40} is a further zoom of the previous lower right panels showing the $Spitzer$ TF sample (green) and adds the TRGB and Cepheid contributions (blue) and SNI$a$ (red).

\begin{figure}[!]
\begin{center} 
\includegraphics[scale=.4]{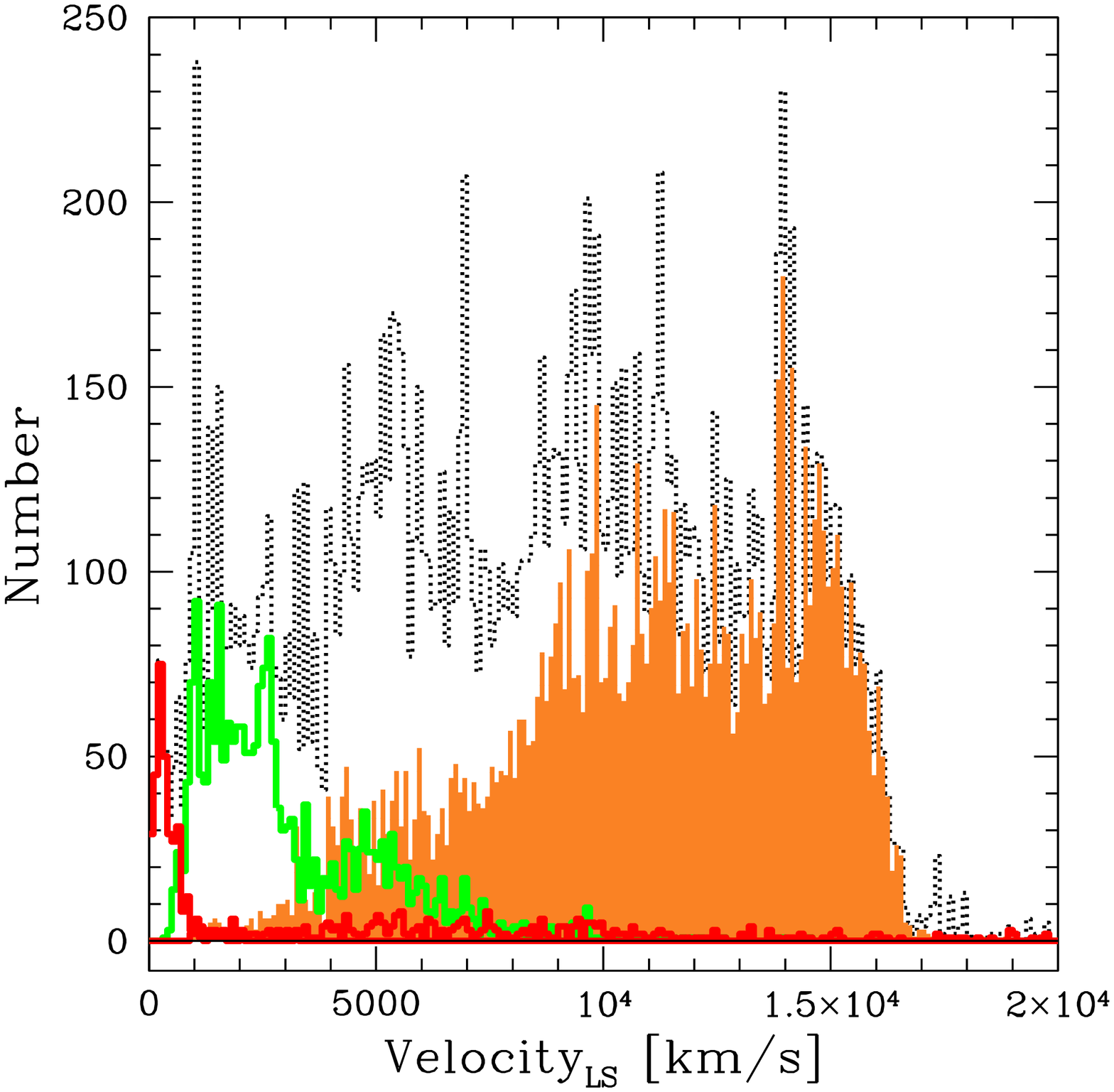}
\caption{Velocity distribution of the sample.  The distribution for all galaxies in CF3 is given by the black dashed histogram.  The 6dFGS component is shown in orange and the $Spitzer$ TF sample is in green.  The red histogram combines the TRGB and Cepheid contributions (the peak at low velocities) and SNI$a$ (strung out over a wide range of velocities).  Velocities are group averaged values to minimize the effects of velocity dispersions in groups.}
\label{histv}
\end{center}
\end{figure}

The most evident feature of the maps is the tremendous numeric contribution of the 6dFGS sample, in orange in the left panels of Figures \ref{xy4} and \ref{xy4zoom100}.  Where there had been a deficiency of information in the celestial south in CF2, now the coverage is dense in that sector.  The distribution of the 6dFGS sample \citep{2014MNRAS.445.2677S} in velocity is seen in the orange histogram of Figure~\ref{histv}.  This sample dominates the global histogram in the range $8,000 - 16,000$~\kms.

The $Spitzer$ TF sample is more local.  It is to be recalled that there are three dominant programatic contributions.  The galaxies from $S^4G$ \citep{2010PASP..122.1397S} are constrained to 40~Mpc ($\sim3000$~\kms) and $\vert b \vert >30^{\circ}$.  Our {\it Cosmic Flows with Spitzer} program \citep{2014MNRAS.444..527S} complemented $S^4G$ by extending to low Galactic latitudes, pushing into the wedges of incompletion evident in Figures \ref{xy4} $-$ \ref{xyzoom40}.  Then this latter program used its remaining available observing time on extreme edge-on disk systems at $3,000 - 10,000$~\kms.

The SNI$a$ sample has grown incrementally from CF2.  Although the numbers are modest (391), the events are dispersed, they explore larger distances, and each contribution has a weight $4-7$ times greater than a TF or FP measure.  The combined Cepheid and TRGB sample has also grown incrementally.  The associated galaxies are nearby, as seen in Figure~\ref{xyzoom40}.  Our understanding of the kinematics of the region within 10~Mpc is now very good \citep{2015ApJ...805..144K}.

\begin{figure}[!]
\begin{center} 
\includegraphics[scale=.95]{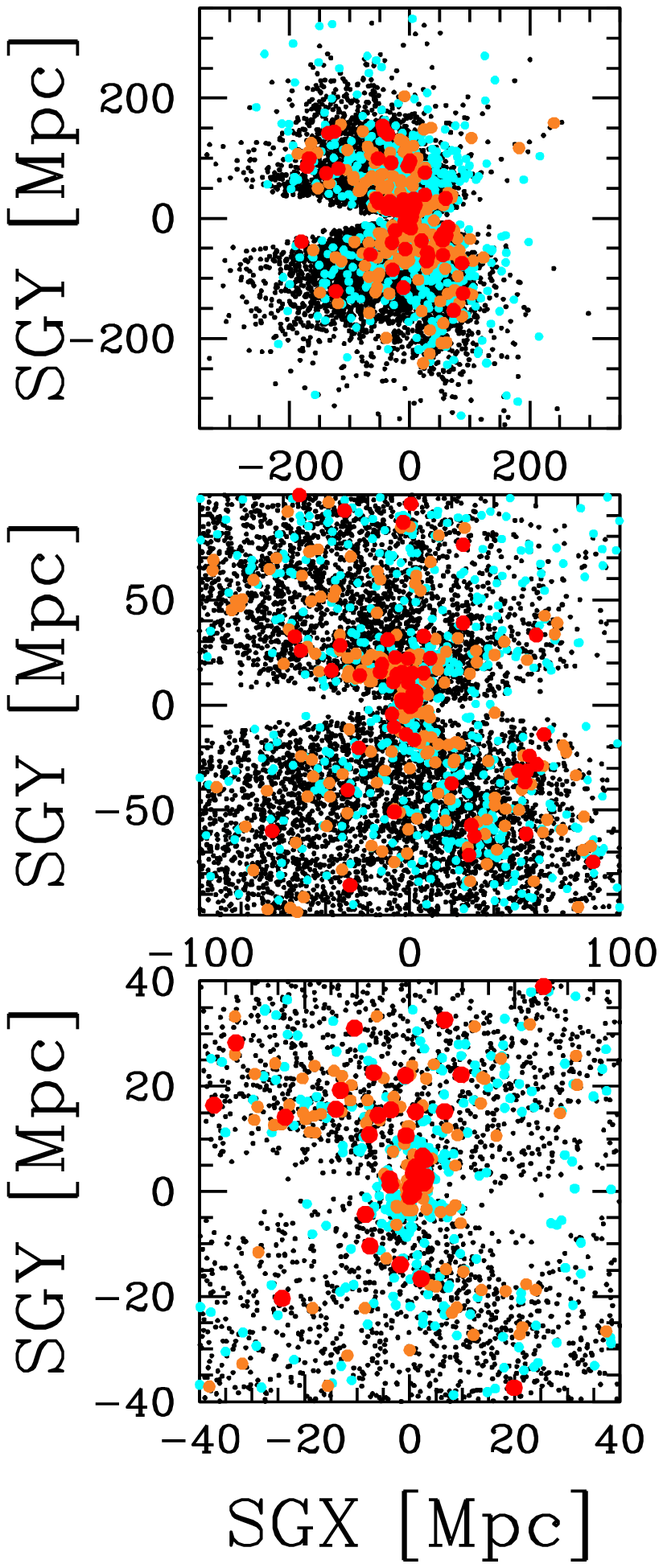}
\caption{Projection of grouped CF3 distances into supergalactic SGX vs. SGY.  Groups with $e_{\mu}^{gp}\le0.10$ are in red, with $0.10<e_{\mu}^{gp}\le0.16$ are in orange, with $0.16<e_{\mu}^{gp}\le0.25$ are in cyan, and with $e_{\mu}^{gp}>0.25$ are black.  The three panels illustrate three different scales.}
\label{xygp3}
\end{center}
\end{figure}

Figure~\ref{xygp3} is a variant of Figures \ref{xy4} $-$ \ref{xyzoom40}.  Here, nests are shown with increasing resolution from the top panel to the bottom.  The symbol colors and sizes distinguish nests by distance uncertainty, strongly correlated with the number of measures.  The information is richest nearby, as would be expected.

{\it Cosmicflows-3} is by far the largest collection of distances at this time.  It will be exciting to translate the distances into deviations from cosmic expansion and develop compatible models of the distribution of matter.  The radial component of a deviation from cosmic expansion, the so-called peculiar velocity is 
\begin{equation}
V_{pec} = (V_{mod} - H_0 d)/(1 + H_0 d/c)
\label{eq:vpds}
\end{equation}
where $V_{mod}$ is the velocity of a galaxy with cosmological modifications as defined in Eq.~\ref{hi} \citep{2014MNRAS.442.1117D}.   As is well known, though, peculiar velocity uncertainties grow with distance and become larger than physical peculiar velocities for individual measurements already at systemic velocities of $2,000 - 3,000$~\kms\ depending on the methodology.  Averaging within groups helps considerably.  Still, further smoothing is inevitably necessary.  Fortunately the task is not hopeless because of large scale coherence in galaxy motions.

 \citet{2014MNRAS.445.2677S} alert us to directionally dependent departures between the velocity field inferred from 6dFGS distances and the velocity fields derived from two independent redshift surveys \citep{1999MNRAS.308....1B, 2006MNRAS.373...45E}.   Their peculiar velocities are more positive than anticipated from models derived from the redshift surveys in the direction toward the Shapley concentration and more negative than anticipated toward the Pisces-Cetus supercluster \citep{1992ApJ...388....9T}.  The implications of these concerns will be evaluated with detailed modeling in later studies.

Another issue of concern is the reliability of our determination of the Hubble Constant of 76.2~\kmsMpc.  This value is slightly higher than we determined recently of 74.4 \citep{2014ApJ...792..129N} and 75.2 \citep{2014MNRAS.444..527S}.  The increase here results from the addition of 26 galaxies to the 13 cluster TF template, minor cluster membership revisions, and some improved inclinations.  The changes are slight but remind us of sensitivity to systematics.

We determine $H_0$ from a zero point calibration of SNI$a$ and then application to two samples of SNI$a$ at $0.05 < z < 0.6$, a range that should not be affected by peculiar velocities nor is strongly dependent on choice of cosmology.  The zero point is anchored by Cepheid and TRGB distances and bootstraps through distances to clusters set by SBF, TF, and the FP programs ENEAR, EFAR, and SMAC,  The 6dFGS material is not used because of limited sample overlap.  An alternative strategy is to jump directly to a SNI$a$ calibration using the few SNI$a$ host galaxies with Cepheid distances \citep{2011ApJ...730..119R}.  It is seen in Figure~\ref{sncal} that these contributions to our calibration lie slightly below the mean.  A calibration based only on these galaxies with Cepheid distances gives a value of $H_0$ 3.8\% lower of 73.4, consistent with Riess et al.  Estimates of systematics should encompass this difference.

{\it Cosmicflows-3} does not directly provide peculiar velocities.  Errors that are Gaussian distributed in distance modulus are lognormal in distance and peculiar velocities, skewing peculiar velocity measurements to negative values.  There are interesting approaches to address this problem \citep{1995MNRAS.276.1391N, 2015MNRAS.450.1868W}.  The great concern is a form of Malmquist bias \citep{1992scma.conf..201L, 1995PhR...261..271S}.  If peculiar velocities are evaluated at the sites of measured distances there will be  artificial flows.  Galaxies will have scattered from the places of physical origin that are  most represented and tend to have erroneous peculiar velocity components that point back to their true positions.  This generic description applies to both the so-called homogeneous and inhomogeneous Malmquist biases.  Biases are much reduced by evaluating distance measures at the sites of systematic velocities since observed velocities have small errors.

We adopt the stance that distances and peculiar velocities should not be confounded.  The effort in this work is to provide distances that have uncertainties but are individually unbiased.  We may not be fully successful, with so much dependence on literature sources and as yet insufficient overlaps in some instances.  In any event, what is provided are distances in Mpc, independent of velocities.  Velocities are provided too, and together with distances one can infer a Hubble Constant, modulo flows and systematics.  

\begin{figure*}[!]
\begin{center} 
\includegraphics[scale=.5]{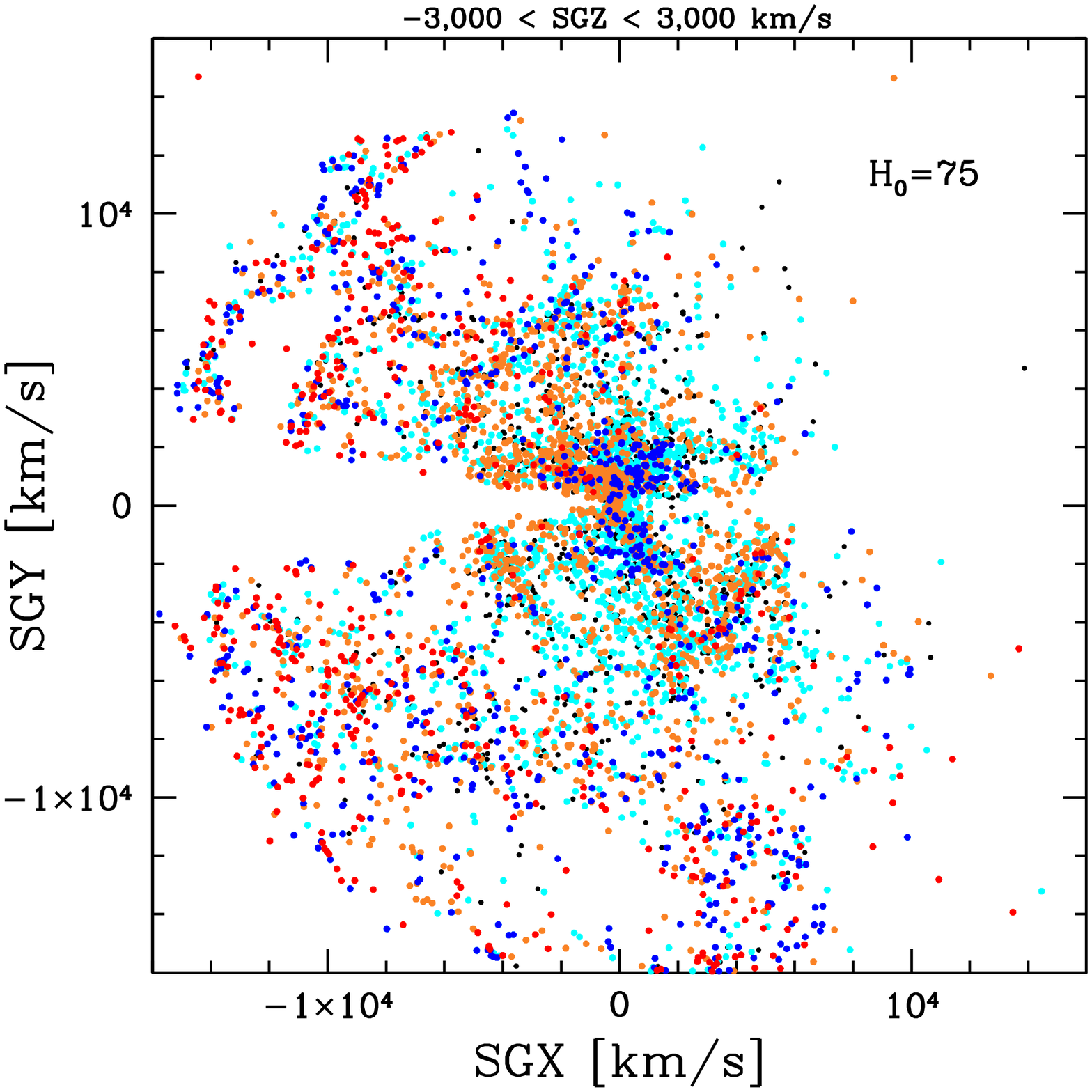}
\includegraphics[scale=.5]{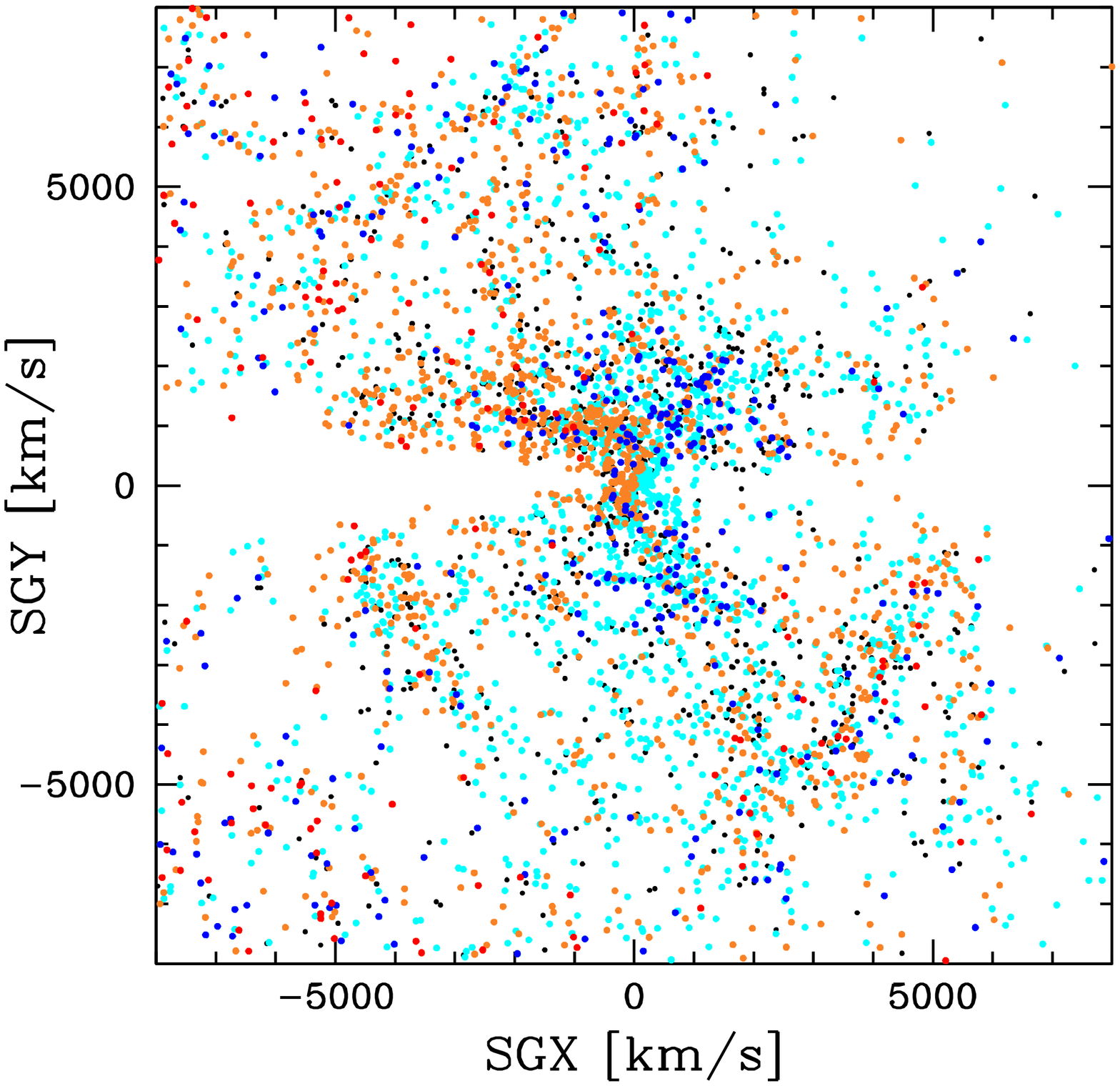}
\caption{Peculiar velocities within $\pm3000$~\kms\ of the supergalactic equator. $V_{pec}^{wf} > 100$~\kms\ orange, $>800$~\kms\ red. $V_{pec}^{wf} < -100$~\kms\ cyan, $<-800$~\kms\ blue. Lower panel is a blow-up of the central region.  $H_0 =75$~\kmsMpc\ assumed.}
\label{xyvp}
\end{center}
\end{figure*}

A first-look visual impression of the peculiar velocity field is given in Figure~\ref{xyvp} with two scales.  Objects within $\pm 3,000$~\kms\ of the supergalactic equator drawn from the groups catalog are plotted at positions given by velocities in the CMB frame with colors indicative of peculiar velocities.  The peculiar velocities represented in the plots are calculated using the formulation by \citet{2015MNRAS.450.1868W} that are statistically unbiased and have Gaussian distributed errors: 
\begin{equation}
V_{pec}^{wf} = {{V_{mod}}\over{1+V_{mod}/c}} {\rm log}(V_{mod}/H_0 d).
\label{eq:vpwf}
\end{equation}
This description of peculiar velocities breaks down nearby where real peculiar velocities are substantial compared with the estimator.  We invoke a ramp within $V_{LS}=3,000$~\kms, linearly transitioning from $V_{pec}$ given by Eq.~\ref{eq:vpds} to $V_{pec}^{wf}$ given by Eq.~\ref{eq:vpwf} as $V_{LS}$ runs from zero to 3,000~\kms. 

\begin{figure}[t]
\begin{center} 
\includegraphics[scale=.42]{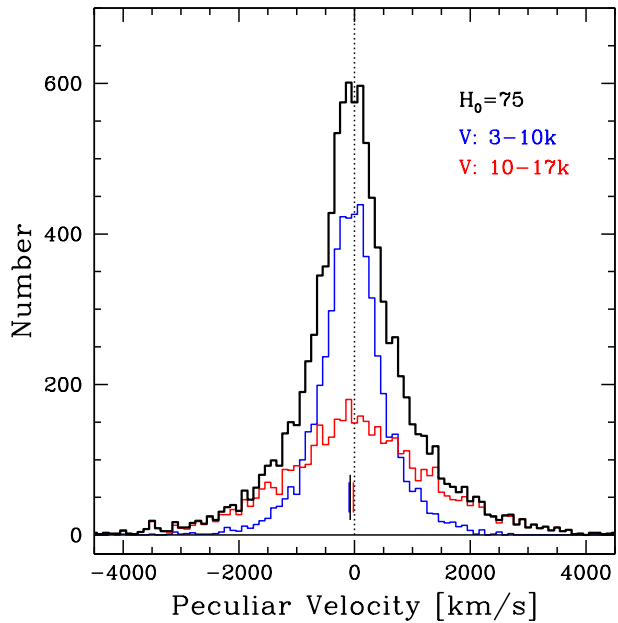}
\caption{Histogram of peculiar velocities, assuming $H_0=75$~\kmsMpc.  The black histogram represents the entire sample.  The blue histogram is built from galaxies with velocities between 3,000 and 10,000~\kms\ and the red histogram, from galaxies outside 10,000~\kms.}
\label{histvp}
\end{center}
\end{figure}

\begin{figure}[t]
\begin{center} 
\includegraphics[scale=.4]{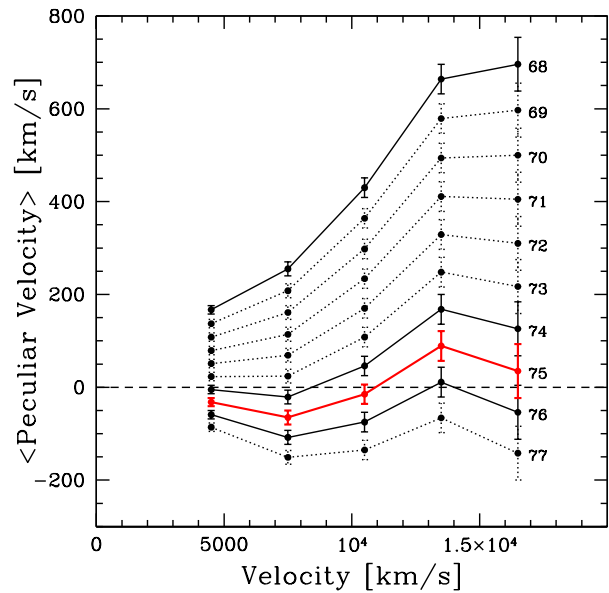}
\caption{Dependence of peculiar velocities on choice of $H_0$.  Mean peculiar velocities and standard deviations are given in systemic velocity bins for choices of $H_0$ between 68 and 77 \kmsMpc.}
\label{hdepend}
\end{center}
\end{figure}

The most evident feature of Figure~\ref{xyvp} is the prominence of blue shades at positive SGX, negative SGY and red shades at negative SGX, positive SGY.  This trend is a manifestation of the well known flow \citep{1988ApJ...326...19L} toward the so-called Great Attractor.  There is great interest in the amplitude and extent of bulk flows \citep{2010MNRAS.407.2328F, 2011ApJ...736...93N, 2015MNRAS.449.4494H}.
Any comment on higher order structure is beyond the scope of this paper.  However we should dwell on the peculiar velocity monopole term: overall expansion or contraction with no preferred direction.  It is seen that there is dependence on $H_0$ with either of Eqs.~\ref{eq:vpds} or \ref{eq:vpwf}.  As the accepted value of $H_0$ is increased, peculiar velocities become more negative, so with a very large $H_0$ there is a general infall of galaxies.  By contrast, as the choice of $H_0$ is decreased, peculiar velocities become more positive, tending toward outflow.  The display seen in Figure~\ref{xyvp} is based on the choice $H_0 = 75$~\kmsMpc.  The justification for the choice is given with Figures~\ref{histvp} and \ref{hdepend}.   The first of these two figures presents histograms of $V_{pec}^{wf}$ assuming $H_0=75$, with separation between galaxies within and beyond 10,000~\kms\  given by the colored histograms and the ensemble in black.  The vertical bars near the origin mark the median values of the three histograms ($-77$, $-96$, and $-20$ \kms\ for the ensemble, inner, and outer samples, respectively).  

Figure~\ref{hdepend} deserves particular attention regarding the choice of $H_0$.  The points with error bars give mean values for peculiar velocities in systemic velocity bins (ignoring $V_{LS} < 3000$~\kms\ where local effects dominated). Dotted and solid lines link bins associated with the same values of $H_0$.  Below the dashed line at zero peculiar velocity is the domain of monopole infall and above the dashed line is the domain of outflow.  A characteristic of a poor choice of $H_0$ would be a monotonic increase in mean peculiar velocities with increasing systemic velocity if $H_0$ is taken too low and the inverse if $H_0$ is taken too high.  From these considerations, a value of $H_0 \sim 75$~\kmsMpc\ is preferred.  The trends in peculiar velocity with systemic velocity only flatten out around $H_0 \sim 76.5$, but there would be a generalized infall of $\sim -80$~\kms\ in that case.  The choice $H_0=75$, shown in red, minimizes the monopole term with the ensemble of the {\it Cosmicflows-3} distances.  With this choice, there is a net tiny infall within 10,000~\kms\ and a net tiny outflow beyond 10,000~\kms.  

We are unable to know if this minimization of the monopole describes the true situation.  If the true value of $H_0$ is 76.2 as indicated by the SNI$a$ calibration then there would be an overall inflow within the region of study.  The implication would be that the local region is over dense which would be a disputed conclusion.   Alternatively, the value of $H_0=68$ from studies of CMB fluctuations \citep{2015arXiv150201589P} can be entertained.  The dramatic implication seen in Figure~\ref{hdepend} would be a massive outflow reaching 700~\kms\ by $0.05c$. The situation with such a low value of $H_0$ is unrealistic.    The displacement of the median peculiar velocity for the ensemble is $+40$~\kms\ as $H_0$ is decreased one unit.  A tolerance of $\pm100$~\kms\ on the monopole averaged over all velocities $<16,000$~\kms\ translates to a restriction on $H_0$ to $75\pm2$.  We can only reconcile with the Planck Collaboration result if there is an unexpectedly large error in the zero point calibration of our distances.
Views on uncertainties in the initial steps of the distance ladder have been expressed by \citet{2014MNRAS.440.1138E} and, as this paper goes to press, by \citet{2016arXiv160401424R}.

\section{Summary}

Distances are assembled for 17,669 galaxies; 7,865 in 1,704 nests with two or more measures and 9,804 singles.  {\it Cosmicflows-3} extends the previous catalog with two major new contributions: our 2,257 distances based on the TF method using [3.6] band {\it Spitzer Space Telescope} photometry \citep{2014MNRAS.444..527S} and 8,885 literature distances based on the FP method from the 6dFGS collaboration \citep{2014MNRAS.443.1231C, 2014MNRAS.445.2677S}.  There are also two incremental extensions: a 29\% increase in the number of TRGB distances arising from our ongoing {\it Hubble Space Telescope} program \citep{2009AJ....138..332J} and a 27\% increase in the number of SNI$a$ distances extracted from the literature \citep{2014ApJ...795...44R, 2015ApJS..219...13W}.

The implementation of a new group catalog \citep{2015AJ....149..171T} is an important supplement.  Group (nest) positions and velocities are averaged across known members and distances are averaged across those members with distance estimates.

A global value of the Hubble Constant of $76.2 \pm 3.4$ (rand.) $\pm 2.7$ (sys.)~\kmsMpc\ is determined by establishing the SNI$a$ zero point with 168 comparisons of distances to nests and nine direct Cepheid comparisons and then using that calibration to establish $H_0$ at $0.05 < z < 0.6$ averaging two SNI$a$ samples  \citep{2010ApJ...716..712A, 2014ApJ...795...44R}.  An alternate estimate of the Hubble Constant is provided by the minimization of the monopole term in the ensemble of peculiar velocities.  If global infall or outflow is assumed to be less than 100~\kms\ then $H_0 = 75 \pm 2$~\kms.

There are as yet few cross-checks with the important 6dFGS component of the collection.  Hopefully the situation will improve with the fourth release of {\it Cosmicflows}, anticipated to include a large all-sky sample of TF distances using $3.4 \mu$m and $4.5\mu$m photometry from WISE, the {\it Wide-field Infrared Satellite Explorer}.  In the meantime, in addition to the on-line resources provided by this journal, {\it Cosmicflows-3} and updates can be accessed at http://edd.ifa.hawaii.edu.

\bigskip\noindent
Note added in proof: Replacing the Cepheid $-$ SNI$a$ comparisons in Fig.~12 with values given in Riess et al. (2016) reduces the $H_0$ calibration illustrated in Fig.~13 from 76.2 to 75.5~\kmsMpc.

\bigskip
\noindent
{\bf Acknowledgements}

Many people have contributed directly or indirectly to this substantial undertaking.  Special thanks to Igor Karachentsev, Dmitry Makarov, Don Neill, Luca Rizzi, Mark Seibert, Ed Shaya, Kartik Sheth, and Po-Feng Wu.  It is hard to imagine how this catalog could have been assembled without the resources of NED, the {\it NASA/IPAC Extragalactic Database}, and the Lyon extragalactic database {\it HyperLeda}. 
Financial support for the Cosmicflows program has been provided by the US National Science Foundation award AST09-08846, an award from the Jet Propulsion Lab for observations with {\it Spitzer Space Telescope}, and NASA award NNX12AE70G for analysis of data from the {\it Wide-field Infrared Survey Explorer}.  Additional support has been provided by the Lyon Institute of Origins under grant ANR-10-LABX-66 and the CNRS under PICS-06233.
  
\bibliography{paper}
\bibliographystyle{aasjournal}

\begin{table*}
\caption{Summary Group Properties \label{grouped}}
\begin{splittabular}{crccrccBrrrrrrrrrrrrrBrrrrrrrrrr}
\hline
 Nest & $N_d^{gp}$ & $<\mu>^{gp}$ & $\epsilon_{\mu}^{gp}$ & $d^{gp}$ & Abell & Gp Name & $N_v$ & PGC1 & $Glon^{gp}$ & $Glat^{gp}$ & $SGL^{gp}$ & $SGB^{gp}$ & $LogL^{gp}$ & $cf$ & $\sigma_p$ & $R_{2t}$ & $V_h^{gp}$ & $V_{gsr}^{gp}$ & $V_{ls}^{gp}$ & $V_{cmb}^{gp}$ & $V_{mod}^{gp}$ & $V_{rms}$ & $M_{12}^{bw}$ & $M_{12}^L$ & LDC & HDC & 2M++ & M\&K & Icnt \\
\hline
  100002 &  161 &  31.01 &  0.02 &   15.9 &        &      Virgo &  197 &    41220 &  284.09 &   74.41 &  103.0008 &   -2.3248 &  12.94 &   1.00 &   707 &  1.920 &   1156 &   1098 &   1064 &   1485 &   1491 &   670 &    935.000 &    705.000 &   852 &   716 &     0 &  5066891 &     1 \\
  200009 &  113 &  36.33 &  0.04 &  184.4 &  A3716 &            &   52 &    66047 &  345.81 &  -39.39 &  224.6483 &   17.2124 &  14.21 &   9.90 &  1950 &  5.295 &  14072 &  14028 &  13998 &  13924 &  14434 &   916 &   4750.000 &  14800.001 &     0 &     0 &  3385 &        0 &  2834 \\
  100010 &  112 &  36.45 &  0.04 &  194.7 &  A3558 &            &   55 &    47202 &  312.07 &   30.45 &  149.1678 &   -1.3439 &  14.18 &   7.95 &  1908 &  5.180 &  14414 &  14263 &  14192 &  14697 &  15265 &  1002 &   4950.000 &  13900.000 &     0 &     0 &  2182 &        0 &  2551 \\
  100001 &  106 &  34.92 &  0.04 &   96.5 &  A1656 &       Coma &  136 &    44715 &   58.99 &   88.14 &   89.6226 &    8.1461 &  13.40 &   1.65 &  1045 &  2.839 &   6926 &   6940 &   6926 &   7195 &   7331 &   886 &   2120.000 &   2280.000 &   890 &   734 &  1952 &        0 &  2493 \\
  100007 &   99 &  35.86 &  0.04 &  148.7 &  A2147 &            &   86 &    56962 &   30.43 &   44.46 &  108.5182 &   49.0878 &  13.77 &   3.30 &  1385 &  3.761 &  11180 &  11283 &  11273 &  11270 &  11603 &  1261 &   7080.000 &   5310.000 &     0 &     0 &  2138 &        0 &  2652 \\
  100006 &   69 &  33.67 &  0.05 &   54.3 &  A1060 &      Hydra &   85 &    31478 &  269.60 &   26.41 &  139.4478 &  -37.6063 &  12.73 &   1.14 &   588 &  1.597 &   3712 &   3490 &   3424 &   4055 &   4099 &   648 &    585.000 &    407.000 &   712 &   592 &  1122 &  5066799 &   487 \\
  100003 &   66 &  33.04 &  0.04 &   40.5 &  A3526 &  Centaurus &  113 &    43296 &  302.20 &   21.74 &  156.2336 &  -11.5868 &  13.00 &   1.12 &   746 &  2.027 &   3550 &   3361 &   3285 &   3834 &   3873 &   822 &   1440.000 &    831.000 &   881 &   722 &  1107 &  5066908 &   441 \\
  200015 &   65 &  31.36 &  0.03 &   18.7 &        &     Fornax &   49 &    12651 &  236.82 &  -54.80 &  262.8089 &  -40.9336 &  12.33 &   1.00 &   414 &  1.123 &   1457 &   1326 &   1319 &   1357 &   1362 &   301 &    107.000 &    141.000 &   391 &   235 &     2 &  5066701 &   843 \\
  100004 &   62 &  35.47 &  0.06 &  124.3 &  A2199 &            &   81 &    58265 &   63.86 &   43.84 &   71.5103 &   49.7851 &  13.41 &   2.26 &  1057 &  2.871 &   9177 &   9348 &   9369 &   9202 &   9424 &   740 &   1880.000 &   2360.000 &  1134 &   936 &  1858 &        0 &  2642 \\
  200007 &   61 &  35.83 &  0.06 &  146.9 &  A0496 &            &   54 &    15524 &  209.07 &  -36.70 &  296.2707 &  -55.3232 &  13.28 &   2.37 &   954 &  2.591 &   9904 &   9794 &   9803 &   9851 &  10106 &   529 &    915.000 &   1730.000 &   307 &   294 &  3653 &        0 &  2181 \\
\hline
\end{splittabular}
\end{table*}

\clearpage
\newpage
\thispagestyle{empty}
\onecolumngrid
\floattable
\begin{sidewaystable*}
\caption{Individual Galaxy Properties \label{individual}}
\fontsize{2}{4}\selectfont
\begin{splittabular}{rrrccccccccccccrcccrrcrccBccrrrrrcrrrrrrrccrccrBclrrrrrrrrrrrrrrrrrrrrrrr}
\hline
PGC  & $d$  & $N_d$  & $<\mu>$  & $\epsilon_{\mu}$  & C  & T  & L  & M  & S  & N  & H  & I  & F  & P  & $\mu_{cf2}$  & $\epsilon_{\mu}^{cf2}$  & SN  & $N_{sn}$  & $\mu_{sn}$  & $\mu_{spit}$  & $\epsilon_{\mu}^{spit}$  & $\mu_{6df}$  & $\epsilon_{\mu}^{6df}$  & $M_t$  & $RAJ$  & $DecJ$  & $Glon$  & $Glat$  & $SGL$  & $SGB$  & $Ty$  & $A_{sf}$  & $B_t$  & $K_s$  & $V_h$  & $V_{gsr}$  & $V_{ls}$  & $V_{cmb}$  & $V_{mod}$  & Name  &  Nest  & $N_d^{gp}$  & $<\mu>^{gp}$  & $\epsilon_{\mu}^{gp}$  & $d^{gp}$  & Abell  & Gp Name  & $N_v$  & PGC1  & $Glon^{gp}$  & $Glat^{gp}$  & $SGL^{gp}$  & $SGB^{gp}$  & $LogL^{gp}$  & $cf$  & $\sigma_p$  & $R_{2t}$  & $V_h^{gp}$  & $V_{gsr}^{gp}$  & $V_{ls}^{gp}$  & $V_{cmb}^{gp}$  & $V_{mod}^{gp}$  & $V_{rms}$  & $M_{12}^{bw}$  & $M_{12}^L$  & LDC  & HDC  & 2M++  & M\&K  & Icnt \\
\hline
      4 &  50.58 & 1 & 33.52 & 0.40 &   &   &   &   &   &   & H &   &   &   & 33.52 & 0.40 &         & 0 &  0.00 &  0.00 & 0.00 &  0.00 & 0.00 &     & 000003.5 & +230515 & 107.8322 & -38.2729 & 316.0587 &  18.4514 &  5.0 & 0.331 & 16.88 &  0.00 &  4458 &  4638 &  4706 &  4109 &  4154 &   AGC331060 & 202766 &   1 & 33.52 & 0.40 &  50.6 &       &           &   2 &     120 & 108.41 & -37.98 & 316.5396 &  18.1559 & 11.32 &  1.15 &  170 & 0.460 &  4353 &  4533 &  4601 &  4005 &  4048 &   25 &           &     9.750 &    0 &    0 &    0 &       0 &    0 \\
     27 & 150.66 & 1 & 35.89 & 0.50 &   &   &   &   &   &   &   &   &   & P &  0.00 & 0.00 &         & 0 &  0.00 &  0.00 & 0.00 & 35.89 & 0.50 & 3.7 & 000023.5 & -065610 &  89.6452 & -66.4507 & 285.8972 &  11.1231 &  2.0 & 0.129 & 16.06 &  0.00 & 11311 & 11405 & 11445 & 10959 & 11275 &             &      0 &   1 & 35.89 & 0.50 & 150.7 &       &           &   1 &      27 &  89.65 & -66.45 & 285.8972 &  11.1231 &       &       &      &       & 11311 & 11405 & 11445 & 10959 & 11275 &      &           &           &    0 &    0 &    0 &       0 &    0 \\
     40 & 116.95 & 1 & 35.34 & 0.50 &   &   &   &   &   &   &   &   &   & P &  0.00 & 0.00 &         & 0 &  0.00 &  0.00 & 0.00 & 35.34 & 0.50 & 3.0 & 000035.6 & -014547 &  95.1366 & -61.8596 & 290.9658 &  12.5924 & -1.0 & 0.132 & 15.11 &  0.00 &  7277 &  7388 &  7434 &  6919 &  7045 &             & 209793 &   1 & 35.34 & 0.50 & 116.9 &       &           &   1 &      40 &  95.14 & -61.86 & 290.9657 &  12.5924 & 10.98 &  1.66 &  126 & 0.341 &  7277 &  7388 &  7434 &  6919 &  7045 &      &           &     3.960 &    0 &    0 &    0 &       0 &    0 \\
     51 & 240.99 & 1 & 36.91 & 0.50 &   &   &   &   &   &   &   &   &   & P &  0.00 & 0.00 &         & 0 &  0.00 &  0.00 & 0.00 & 36.91 & 0.50 & 2.8 & 000035.8 & -403432 & 337.6665 & -72.9440 & 254.0322 &  -0.2359 &  0.0 & 0.037 & 16.93 &  0.00 & 15015 & 14983 & 14980 & 14771 & 15342 & PGC000051   & 201987 &   1 & 36.91 & 0.50 & 241.0 &       &           &   2 &     142 & 337.01 & -72.84 & 253.8705 &  -0.3818 & 12.55 &  8.84 &  502 & 1.364 & 14986 & 14953 & 14950 & 14743 & 15315 &  132 &           &   253.000 &    0 &    0 &    0 &       0 &    0 \\
     55 &  73.79 & 1 & 34.34 & 0.40 &   &   &   &   &   &   & H &   &   &   & 34.34 & 0.40 &         & 0 &  0.00 &  0.00 & 0.00 &  0.00 & 0.00 &     & 000037.4 & +333603 & 110.9496 & -28.0856 & 327.0998 &  19.7763 &  5.9 & 0.183 & 17.04 &  0.00 &  4779 &  4979 &  5052 &  4454 &  4507 &    UGC12898 &      0 &   1 & 34.34 & 0.40 &  73.8 &       &           &   1 &      55 & 110.95 & -28.09 & 327.0998 &  19.7763 &       &       &      &       &  4779 &  4979 &  5052 &  4454 &  4507 &      &           &           &    0 &    0 &    0 &       0 &    0 \\
     64 & 211.84 & 1 & 36.63 & 0.50 &   &   &   &   &   &   &   &   &   & P &  0.00 & 0.00 &         & 0 &  0.00 &  0.00 & 0.00 & 36.63 & 0.50 & 0.0 & 000052.3 & -355037 & 350.7982 & -76.1593 & 258.4801 &   1.3810 & -5.0 & 0.042 & 15.53 &  0.00 & 15596 & 15582 & 15585 & 15331 & 15946 &             & 200033 &  11 & 36.25 & 0.15 & 178.1 & A2717 &           &  25 &   72642 & 356.24 & -75.69 & 258.8940 &   2.7238 & 14.10 &  8.84 & 1788 & 4.857 & 15000 & 14991 & 14996 & 14729 & 15300 & 1582 & 17100.000 & 11400.000 &    0 &    0 & 3084 &       0 &    0 \\
     66 & 182.81 & 1 & 36.31 & 0.50 &   &   &   &   &   &   &   &   &   & P &  0.00 & 0.00 &         & 0 &  0.00 &  0.00 & 0.00 & 36.31 & 0.50 & 0.0 & 000053.2 & -355911 & 350.3084 & -76.0782 & 258.3474 &   1.3282 & -5.0 & 0.046 & 15.80 &  0.00 & 14996 & 14981 & 14985 & 14731 & 15300 &             & 200033 &  11 & 36.25 & 0.15 & 178.1 & A2717 &           &  25 &   72642 & 356.24 & -75.69 & 258.8940 &   2.7238 & 14.10 &  8.84 & 1788 & 4.857 & 15000 & 14991 & 14996 & 14729 & 15300 & 1582 & 17100.000 & 11400.000 &    0 &    0 & 3084 &       0 &    0 \\
     70 & 117.49 & 1 & 35.35 & 0.40 &   &   &   &   &   &   & H &   &   &   & 35.35 & 0.40 &         & 0 &  0.00 &  0.00 & 0.00 &  0.00 & 0.00 &     & 000056.1 & +202016 & 107.1780 & -40.9837 & 313.2487 &  17.7662 &  5.8 & 0.282 & 15.61 & 11.26 &  6800 &  6974 &  7040 &  6447 &  6557 &    UGC12900 & 209949 &   1 & 35.35 & 0.40 & 117.5 &       &           &   1 &      70 & 107.18 & -40.98 & 313.2488 &  17.7663 & 10.95 &  1.53 &  122 & 0.332 &  6800 &  6974 &  7040 &  6447 &  6557 &      &           &     3.650 &    0 &    0 &    0 &       0 &    0 \\
     76 &  97.72 & 1 & 34.95 & 0.40 &   &   &   &   &   &   & H &   &   &   & 34.95 & 0.40 &         & 0 &  0.00 &  0.00 & 0.00 &  0.00 & 0.00 &     & 000058.9 & +285441 & 109.8059 & -32.6707 & 322.1728 &  19.1316 &  3.0 & 0.178 & 14.82 & 11.01 &  6920 &  7112 &  7183 &  6583 &  6698 &    UGC12901 & 209247 &   1 & 34.95 & 0.40 &  97.7 &       &           &   1 &      76 & 109.81 & -32.67 & 322.1729 &  19.1316 & 11.09 &  1.54 &  139 & 0.376 &  6920 &  7112 &  7183 &  6583 &  6698 &      &           &     5.310 & 1536 &    0 &    0 &       0 &    0 \\ 
     82 & 114.82 & 1 & 35.30 & 0.50 &   &   &   &   &   &   &   &   &   & P &  0.00 & 0.00 &         & 0 &  0.00 &  0.00 & 0.00 & 35.30 & 0.50 & 0.8 & 000106.0 & -535931 & 318.5468 & -61.5837 & 241.4720 &  -4.9970 & -1.3 & 0.050 & 14.67 &  0.00 &  9426 &  9344 &  9324 &  9251 &  9476 &             & 200218 &   9 & 36.21 & 0.17 & 174.2 &       &           &   7 &     211 & 317.61 & -61.47 & 241.3354 &  -5.4383 & 12.52 &  2.84 &  489 & 1.328 & 10418 & 10334 & 10313 & 10245 & 10520 &  635 &   832.000 &   233.000 &    5 &    0 & 3377 &       0 &    0 \\
\end{splittabular}
\end{sidewaystable*}

\end{document}